\documentclass[usenatbib]{mn2e}

\usepackage{times}
\usepackage{amssymb}
\usepackage{rotating}
\usepackage{xspace}
\usepackage{caption}
\usepackage{amsmath}
\input{epsf}

\title[Reaching the Peak of the Quasar Spectral Energy Distribution I: Observations and Models]
{Reaching the Peak of the Quasar Spectral Energy Distribution I: Observations and Models}

\author[James\,S. Collinson et al.]{James S. Collinson$^1$\thanks{Email: j.s.collinson@durham.ac.uk}, Martin J. Ward$^1$, Chris Done$^{1}$, Hermine Landt$^1$, Martin Elvis$^2$ \newauthor and Jonathan C. McDowell$^2$\\
 \\
$^1$Department of Physics, University of Durham, South Road, Durham, DH1 3LE, UK\\
$^2$Harvard-Smithsonian Center for Astrophysics, 60 Garden St., Cambridge, MA 02138, USA\\
}

\pagerange{\pageref{firstpage}--\pageref{lastpage}} \pubyear{2014}

\newcommand{\xmm}{{\it XMM}\xspace}
\newcommand{\xmmn}{{\it XMM-Newton}\xspace}
\newcommand{\nustar}{{\it NuSTAR}\xspace}

\newcommand{\galex}{{\it GALEX}\xspace}
\newcommand{\gemini}{{\it Gemini}\xspace}

\newcommand{\optxagnf}{{\sc optxagnf}\xspace}
\newcommand{\wabs}{{\sc wabs}\xspace}

\newcommand{\zwabs}{{\sc zwabs}\xspace}
\newcommand{\zredden}{{\sc zredden}\xspace}
\newcommand{\xspec}{{\sc xspec}\xspace}
\newcommand{\iraf}{{\sc iraf}\xspace}

\def\Lbol{$L_{\rm bol}$}
\def\LEdd{$L_{\rm Edd}$}

\def\risco{$r_{\rm isco}$\xspace}

\def\rcor{$r_{\rm cor}$\xspace}
\def\fcol{$f_{\rm col}$\xspace}
\def\fsx{$f_{\rm SX}$\xspace}
\def\mdot{{\it \.m}\xspace}
\def\M_BH{$M_{\rm BH}$\xspace}

\def\rchi{{$\chi^{2}_{\rm red}$}}

\def\H0{{\rm ~km~s^{-1}~Mpc^{-1}}}

\def\eg{{e.g.\ }}

\def\la{\mathrel{\hbox{\rlap{\hbox{\lower4pt\hbox{$\sim$}}}{\raise2pt\hbox{$<$}}
}}}
\def\ga{\mathrel{\hbox{\rlap{\hbox{\lower4pt\hbox{$\sim$}}}{\raise2pt\hbox{$>$}}
}}}
\def\d25{$D_{25}$}
\def\nh{{$N_{\rm H}$}}
\def\Ha{{H$\alpha$}\xspace}
\def\Hb{{H$\beta$}\xspace}
\def\Hg{{H$\gamma$}\xspace}

\def\Civ{{C$\,${\sc iv}}\xspace}
\def\Mgii{{Mg$\,${\sc ii}}\xspace}
\def\Oiii{{[O$\,${\sc iii}]}\xspace}
\def\Feii{{Fe$\,${\sc ii}}\xspace}
\def\Sii{{[S$\,${\sc ii}]}\xspace}
\def\Nii{{[N$\,${\sc ii}]}\xspace}

\begin{document}

\maketitle

\label{firstpage}

\begin{abstract}

We perform a spectral analysis of a sample of 11 medium redshift ($1.5 \lesssim z \lesssim 2.2$) quasars. Our sample all have optical spectra from the SDSS, infrared spectra from GNIRS and TSPEC, and X-ray spectra from \xmmn . We first analyse the Balmer broad emission line profiles which are shifted into the IR spectra to constrain black hole masses. Then we fit an energy-conserving, three component accretion model of the broadband spectral energy distribution (SED) to our multiwavelength data. Five out of the 11 quasars show evidence of an SED peak, allowing us to constrain their bolometric luminosity from these models and estimate their mass accretion rates. Based on our limited sample, we suggest that estimating bolometric luminosities from $L_{5100 \rm \mathring{A}}$ and $L_{\rm 2-10\,keV}$ may be unreliable, as has been also noted for a low-redshift, X-ray selected AGN sample.

\end{abstract}


\section{Introduction} \label{sec:introduction}
\subsection{Background} \label{subsec:bg}
Multi-wavelength studies of quasars are an important means to further our understanding of these objects. Interpreting the quasar spectral energy distribution (SED) remains a challenge in astrophysics, although the last few decades have seen significant advances in the field (\eg \citealt{ward87}, \citealt{elvis94}, \citealt{vasudevan07}). It is now widely accepted that accretion of gas onto a central supermassive black hole (SMBH) is the ultimate source powering these extremely luminous objects. The SED contains clues to the geometry and properties of the matter in the regions close to the black hole (BH). Better understanding of the SEDs also has consequences for cosmology, as there is strong evidence that galaxy formation is influenced by the quasar (hereafter referred to as an active galactic nucleus, AGN) through the process of feedback (\eg \citealt{mccarthy10}).

The simplest interpretation of the unified model of AGN proposes that most differences in the observed SED properties can be attributed to differences in the orientation of the accretion flow and line-of-sight absorption. Those AGN characterised by a direct sightline towards the central engine are classed as Type I, and those with optically thick material along the line-of-sight are classified as Type II  \citep{antonucci93}. In practice, however, this simple picture is incomplete, and other physical differences in the BH itself (\eg spin) and the properties of the infalling matter influence the observed SED (\citealt{boroson92}, \citealt{done12}).

This situation has motivated us to produce new broadband SED models, based on theoretical considerations and on empirical studies of accreting stellar-mass BHs found in X-ray binaries. Combining AGN SED models with representations of galactic and extragalactic extinction, both via dust and photoelectic absorption, enables fitting of the multi-wavelength data from an AGN, and recovery of the intrinsic SED, which relates directly to the properties of the BH and the material it accretes. Previous studies by \nocite{jin12_1} \nocite{jin12_2} \nocite{jin12_3} Jin et al.\ (2012a, b, c), hereafter J12a, b, c (collectively J12) have successfully employed this technique. The peak of the energy output occurs at far-ultraviolet/ultra soft X-ray energies in the majority of these low redshift AGN. These energy ranges are mostly unobservable due to photoelectric absorption by neutral Hydrogen in the inter-galactic medium along the line-of-sight to the source (the so-called Gunn-Peterson trough, \citealt{gunn65}), and that intrinsic to the Milky Way.

In order to overcome this restriction we consider higher redshift AGN, in which it is expected that the peak of the SED would be shifted into the observable optical/near-UV energy range. This occurs for two reasons; the first is simply the redshift, and the second is because these more distant, luminous AGN contain more massive BHs which have cooler accretion discs peaking at lower energies (\citealt{mclure04}, \citealt{done12}).

In this first paper in a series, we present the sample and discuss the selection process and the data that has been assembled. We then present model fits from rest-frame optical to hard X-rays, and explore the modifying effects of extinction and the presence of a stellar component in the host galaxy. In the next paper (Collinson et al.\ (in prep), hereafter Paper II), we will investigate the parameter space further, including the toroidal dust component. In Paper III, we will apply these findings to the analysis of a larger, statistically significant sample.

\subsection{A Refined AGN SED Model} \label{sec:model}

\begin{figure}
\centering
\includegraphics[width = 0.5\textwidth]{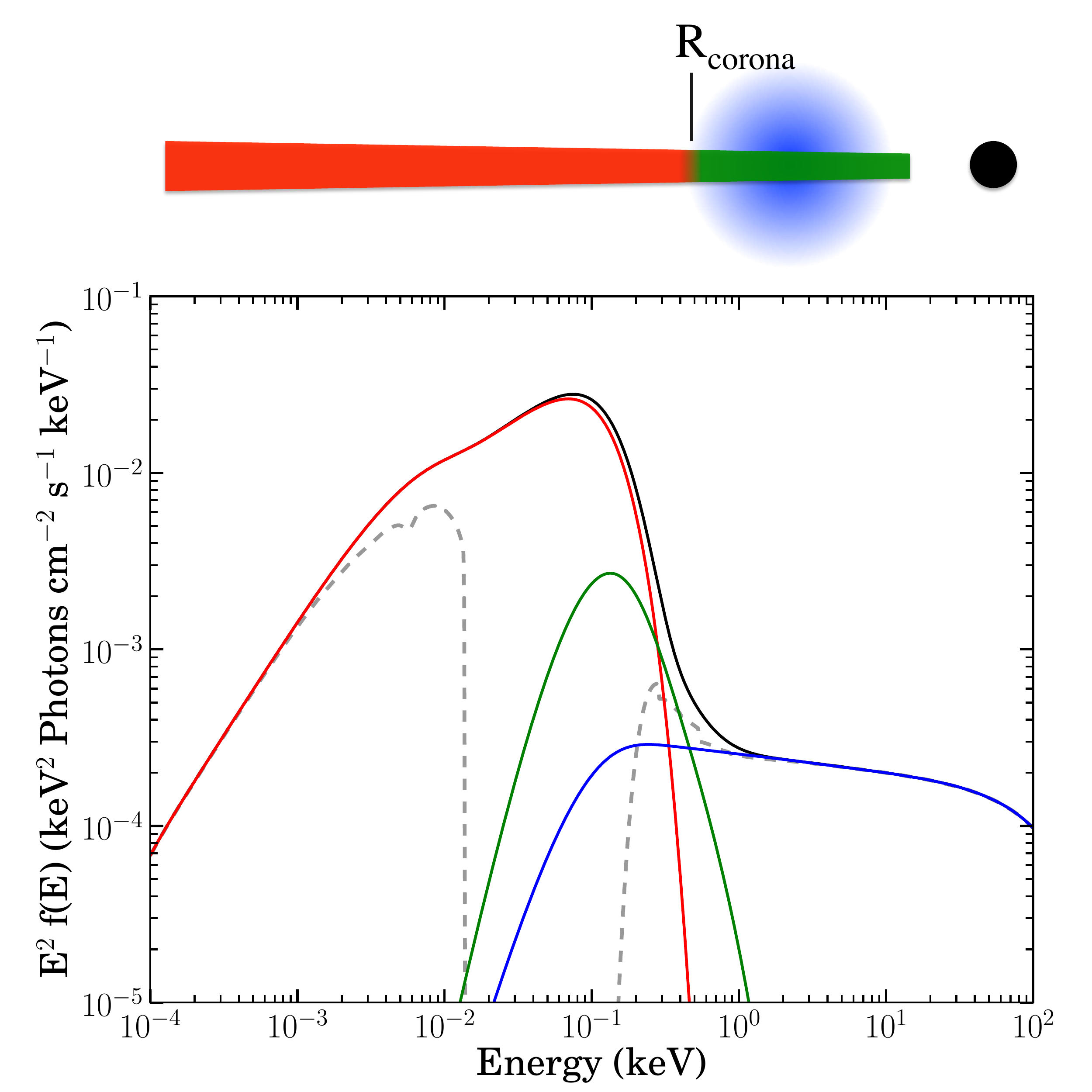}
\caption{\small An SED characteristic of a typical disc-dominated AGN. Colours correspond to the regions shown in the AGN schematic (top), with red representing the AD, green the SX and blue the PLT from the corona. The grey dotted line shows the resultant observed SED when it has been attenuated by typical galactic and extra-galactic extinction and absorption. Figure adapted from Done et al.\ (2012).}
\label{fig:sed}
\end{figure}

We use a multi-component model based on studies of black hole binaries (BHBs) and nearby Narrow-line Seyfert 1s (NLS1s). This model is described in \cite{done12}. It is characterised by three principal components: an accretion disc (AD), a power-law tail (PLT) and a soft X-ray excess (SX). A schematic SED diagram is shown in Fig.\ \ref{fig:sed}. The AD is modelled as a relativistic, geometrically thin, optically thick disc, with each radius in the disc radiating as a blackbody (\eg \citealt{shakura73}). Our latest disc models include a colour correction (\fcol) to account for the fact that the disc is not completely thermalised at all radii. There is also a large contribution from the PLT at high energies, arising from inverse Compton scattering of AD photons by a hot, optically thin corona \citep{zdziarski95}.

The origin of the SX is less well understood, and is the subject of debate. Some postulate that the SX is produced by reflection of hard coronal X-rays off the AD (\eg \citealt{crummy06}, \citealt{fabian09}, \citealt{zoghbi11}), whilst others attribute it to the presence of intervening matter, which complicates the observed emission (\eg \citealt{miller10}), and that a warm, optically thick Comptonised component of the inner AD better describes the observations (\eg \citealt{alston14}, \citealt{gardner14}). \cite{jin13} and \cite{matt14} used long observations with \xmmn and \nustar to test models of the SX in two different AGN, and also found Comptonisation of inner disc photons to better describe their observations. Whilst the model we adopt in this paper assumes the latter origin for the SX, in practice limitations in the quality and energy coverage of our data mean that uncertainties in the origin of the SX cannot be unambiguously resolved.  Additional factors arising from absorption and reflection could be included. The interplay between the model components is complex, but by fitting such physically motivated models to data, we can infer information about the accretion flow properties (\eg \citealt{elvis94}, \citealt{elvis12}).

Importantly, the \cite{done12} AGN SED model we use -- \optxagnf \ -- applies the constraint of energy conservation, as energy output is dependent on the amount of matter accreting on to the BH. This model has so far been tested on a moderate number of AGN (\eg J12, \citealt{done13}, \citealt{matt14}). It has been designed for implementation into NASA's High-Energy Astrophysics Science Archive Research Center (HEASARC) \xspec spectral fitting package.

\subsection{Previous Work} \label{subsec:previous}
Initial studies of AGN focused primarily on ultraviolet (UV), optical and infrared (IR) spectra (\eg \citealt{wills85}, \citealt{zheng97}), but with improvements in ground-based instrumentation and new satellites, it is now routinely possible to study AGN samples across multiple wavelength bands.

\cite{puchnarewicz92} presented a study of optical and X-ray data of 53 AGN with ultra-soft X-ray excesses, and found a bias in their emission line profiles towards narrow linewidths, which had implications for the position and size of the broad line region (BLR). \cite{grupe98} and \cite{grupe99} confirmed this result and also reinforced the findings of \cite{walter93} that there is a Big Blue Bump (BBB) from optical to X-ray spectra.

More recently, J12 presented a medium-sized SED modelling study of 51 AGN. In their study, they assembled optical and X-ray spectra from the Sloan Digital Sky Survey (SDSS) and the European Space Agency (ESA) \xmmn , respectively. This was supplemented, when available, by photometric UV data from the \xmm Optical Monitor (OM). These data ranges are not contiguous and leave the SED peak, unobservable due to the aforementioned absorption, devoid of data. Whilst SDSS and \xmm OM data constrain only the optical edge of the AD, \xmm EPIC data only lies in the energy range of the SX and PLT. Thus the AD peak (and hence SED peak in the disc-dominated objects) was unsampled and information as to its position and shape had to be inferred from the fitted models. Since the AD peak originates from the innermost part of the accretion flow, it contains key information about the BH spin and mass accretion rate, \mdot , and hence the bolometric luminosity of the AGN.

BHs have `no-hair', meaning they are characterised by just three properties -- mass, spin and charge. Much study has been devoted to measuring the mass of a BH (see \S \ref{sec:masstheory}), and charge is negligible in an astrophysical context. There is currently much interest in constraining spin, which can be estimated if some measure of the radius of innermost stable circular orbit (\risco) can be made. In practice this is difficult; \cite{fabian09} and \cite{risaliti13} claim to make such measures from observations of the broad iron K emission line, but this is still controversial (\eg \citealt{miller13}).

\cite{davis11}, \cite{done13}, \cite{matt14} and \cite{capellupo14} have all investigated the potential for using SED modelling to constrain BH spin. \cite{done13} utilised the same energy conserving model discussed in \S \ref{sec:model} to rule out a high-spin solution for the SMBH at the centre of PG1244$+$026. \cite{matt14} also used this model in their study of Ark 120, inferring an intermediate BH spin from simultaneous \xmmn and \nustar observations. \cite{capellupo14} fitted a thin accretion disc model \citep{slone12} to the IR/optical continua of 30 quasars in a similar redshift range to our sample. Using this model they inferred a range of BH spin parameters in their sample.

Whilst previous works have successfully found evidence for the AD turnover by recovering the continuum from high-resolution UV spectra of high-redshift AGN, (\eg \citealt{zheng97}, \citealt{finn14}) no sizeable sample exists with X-ray spectra. Adding to this is the need for reliable BH mass estimates.

\begin{table*}
\caption{\small The names, positions, SDSS pipeline redshifts and UV/IR data sources for the sample of 11 objects. The UV photometry is of limited use for our purposes (see \S \ref{subsec:sedfit}).}
\small
\centering
\begin{tabular}{cccccccc}
\hline
ID & Name & R.A. (J2000) & Decl. (J2000) & Common Name & $z_{\rm SDSS}$ & IR Source & UV Phot$^a$ \\
\hline
1  & J0041$-$0947  & 00 41 49.64 & $-$09 47 05.0  &  $-$     & 1.629 & TSPEC$^b$  & N \\
2  & J0043$+$0114  & 00 43 15.08 & $+$01 14 45.6  &  $-$     & 1.563 & GNIRS 13B  & N \\
3  & J0118$-$0052  & 01 18 27.98 & $-$00 52 39.8  &  QSO B0115$-$0108 & 2.188 & GNIRS 04B & $-$\\
4  & J0157$-$0048  & 01 57 33.87 & $-$00 48 24.4  &  QSO J0157$-$0048 & 1.551 & TSPEC$^b$ & N M2 W1 \\
5  & J0839$+$5754  & 08 39 06.53 & $+$57 54 17.0  &  3C 205  & 1.534 & GNIRS 13B  & N W1 \\
6  & J1021$+$1315  & 10 21 17.74 & $+$13 15 45.9  & $-$     & 1.565 & GNIRS 13B  & W1 B \\
7  & J1044$+$2128  & 10 44 01.13 & $+$21 28 03.9  & $-$     & 1.494 & GNIRS 13B  & N \\
8  & J1240$+$4740  & 12 40 06.70 & $+$47 40 03.3  & $-$     & 1.561 & TSPEC$^b$  & N F U \\
9  & J1350$+$2652  & 13 50 23.68 & $+$26 52 43.1  & QSO B1348$+$2707 & 1.624 & TSPEC$^b$ & N F \\
10 & J2328$+$1500  & 23 28 10.56 & $+$15 00 12.8  & $-$     & 1.536 & GNIRS 13B  & N \\
11 & J2332$+$0000  & 23 32 28.21 & $+$00 00 32.8  & $-$     & 1.604 & GNIRS 13B  & N \\

\hline

\multicolumn{7}{l}{$^a$N $=\,${\it NUV}$_{\rm GALEX}$, F $=\,${\it FUV}$_{\rm GALEX}$, M2 $=\,${\it UVM2}$_{\rm XMM \; OM}$, W1 $=\,${\it UVW1}$_{\rm XMM \; OM}$, B $=\,${\it B}$_{\rm XMM \; OM}$.}\\
\multicolumn{7}{l}{$^b$IR spectral data from TSPEC courtesy of Yue Shen \citep{shen12}.}\\

\end{tabular}
\label{tab:sample}
\end{table*}

\subsection{Black Hole Mass Estimates} \label{sec:masstheory}
Reverberation Mapping (RM), proposed by \cite{blandford82} and employed by \eg \cite{peterson04}, \cite{bentz09} and \cite{denney10}, amongst others, is a technique for accurately determining the mass of BHs in AGN. RM involves measuring the delay in time for variations in the central source intensity (continuum) to reach the broad line region (broad emission lines), and using this as a proxy for the light travel time across the BLR, which is well-correlated with the BH mass. RM requires extensive long-term monitoring programs, and is thus very expensive observationally, however, it has allowed the calibration of BH mass estimates from single-epoch observations of the broad emission lines in AGN optical spectra (\eg \citealt{woo02}, \citealt{greene05}, \citealt{matsuoka13}). The abundance of suitable single epoch spectra means that this method has now been applied to a great many AGN, generally utilising the well-studied Balmer-series Hydrogen lines \Ha and \Hb .

The problem with this technique arises in high-redshift AGN, where the Balmer lines are redshifted to infrared wavelengths, and rest-frame UV lines are shifted to the optical regime. Much effort has been devoted to the use of the rest-UV lines for the purpose of BH mass estimation, but the subject remains contentious. \cite{vestergaard06} studied the scaling relationships between mass estimates from \Hb , \Ha and the rest-UV lines \Civ and \Mgii , finding that \Civ and \Mgii offer viable alternatives to the Balmer lines for this purpose in a sample of low-redshift AGN. However, two studies of high-redshift AGN, \cite{netzer07} and \cite{shen12}, both disagree with this result, contending that the line profile of \Civ is not suitable for mass estimates, but agreeing that \Mgii shows reasonable correlation. Possible explanations put forward for this discrepancy in \Civ are the presence of outflows that influence line profiles in high-redshift, high-Eddington-ratio AGN. It has also been claimed \citep{denney13} that high signal-to-noise ratio (S/N) spectra are required to derive reliable mass estimates from \Civ . In any case, \Hb remains the mass estimator of choice, \citep{woo02} simply by its virtue as the line best calibrated by RM. \cite{greene05} propose an alternative in \Ha which shows excellent correlation with \Hb , but which, due to greater line strength, generally offers better S/N for the line analysis.

\begin{table*} 
\caption{\small Optical/IR spectral observation dates for our sample of 11 objects. S/N values are approximate.}
\small
\centering
\begin{tabular}{ccccccccc}
\hline
ID & Name & Survey & MJD-Plate-Fibre & Optical & Avg. & IR     & IR      & Avg.  \\
   &      &        &                 & Obs. UT & S/N  & Source & Obs. UT & S/N    \\
\hline \hline
1  & J0041$-$0947 & SDSS & 52162-655-172   & 2001-09-10 & 36 & TSPEC    &  2010-01-02 \& 11-28 & 7  \\
\hline
2  & J0043$+$0114 & SDSS & 51794-393-419   & 2000-09-07 & 18 & GNIRS 13B & 2013-08-16 &           12 \\
   &              & BOSS & \bf{55186-3589-707} & 2009-12-21 & 30 &      &  &                         \\
   &              & BOSS & 55444-4222-902  & 2010-09-05 & 33 &          &  &                         \\
\hline
3  & J0118$-$0052 & SDSS & 51789-398-211   & 2000-09-02 & 16 & GNIRS 04B & 2004-11-29 &           9   \\
\hline
4  & J0157$-$0048 & SDSS & 51871-403-213 & 2000-11-23 & 17 & TSPEC  & 2009-11-07 \& 11-28 &  3   \\
   &              & SDSS & 52179-701-294   & 2001-09-27 & 14 &           &  &                         \\
   &              & BOSS & \bf{55449-4233-152}  & 2010-09-10 & 24 &           &  &                         \\
\hline
5  & J0839$+$5754 & SDSS & 54425-1784-495  & 2007-11-21 & 32 & GNIRS 13B & 2013-10-27 &           19  \\
\hline
6  & J1021$+$1315 & SDSS & 53062-1746-491  & 2004-02-27 & 14 & GNIRS 13B & 2014-03-21 &           15  \\
\hline
7  & J1044$+$2128 & SDSS & 54097-2478-411  & 2006-12-28 & 18 & GNIRS 13B & 2014-03-20 &           14  \\
   &              & BOSS & \bf{56039-5874-970} & 2012-04-22 & 28 &       &  &                         \\
\hline
8  & J1240$+$4740 & SDSS & 53089-1455-424  & 2004-03-25 & 16 & TSPEC     & 2011-02-22 &           8  \\
\hline
9  & J1350$+$2652 & SDSS & 53848-2114-105  & 2006-04-23 & 27 & TSPEC     & 2011-02-22 &           8  \\
   &              & BOSS & \bf{56105-6006-260} & 2012-06-27 & 47 &       &  &                        \\
\hline
10 & J2328$+$1500 & SDSS & 52238-746-463   & 2001-11-25 & 10 & GNIRS 13B & 2013-08-18 &           11 \\
\hline
11 & J2332$+$0000 & SDSS & \bf{51821-384-438} & 2000-10-04 & 13 & GNIRS 13B & 2013-08-19 &           13 \\
   &              & SDSS & 52199-681-543   & 2001-10-17 & 13 &           &  &                        \\
   &              & SDSS & 52525-682-355   & 2002-09-08 & 11 &           &  &                        \\

\hline

\end{tabular}
\label{tab:dates}
\end{table*}

\section{Sample and Data Assembly} \label{sec:sample}
\subsection{Sample Selection} \label{subsec:sample}
In order to define an AGN sample in which the SED peak would be observable, we required objects with redshifts around $z$~$\sim$~$2$. For SED modelling, we require data lying on both sides of the UV/soft X-ray absorption trough, as was the case in the J12 sample. We thus required optical spectra, available from the SDSS ($\sim$~$35\%$ sky coverage) and X-ray data, available from \xmmn 's 3XMM DR4 data catalogue ($\sim$~$2\%$ sky coverage). In order to constrain physical parameters for the SED model we also required IR spectral data, so that BH mass estimates from the Balmer lines could be made.

We started by searching the \cite{schneider10} SDSS DR7 QSO catalog for all AGN meeting the following criteria:

\begin{itemize}
\item[$\bullet$]{1.49~$<$~$z$~$<$~1.61: This was so that \Ha \ and \Hb \ would lie in the NIR $H$ and $J$ bands respectively, and \Mgii and \Civ lines would be visible in the SDSS spectra, for comparison purposes.}
\item[$\bullet$]{$K_{\rm 2MASS}$~$<$~16.5: Since we required medium resolution IR spectra, we needed the objects to be suitably bright in the IR bands. Where {\it K} was unavailable in 2MASS, similar constraints were applied to the {\it J} and {\it H} bands. There were 1797 matching AGN after this step.}
\item[$\bullet$]{\xmmn data: We also required X-ray spectra from ESA's \xmmn instrument. 63 objects had matching observations with 18 of these being bright enough to have had a spectrum extracted by the \xmm pipeline (typically requiring $\gtrsim 200$ counts).}
\end{itemize}

We were awarded observing time in Cycle 2013B to use the \gemini GNIRS instrument, sited at Mauna Kea Observatory, Hawai'i, to obtain high quality IR spectra in the $J$, $H$ and $K$ bands. In practice 9 of our objects were visible to GNIRS in 2013B, and we proposed to observe the 6 with the highest SDSS S/N.

Simultaneously, we identified 4 objects in the \cite{shen12} sample with \xmm X-ray data of varying quality, and IR spectral data from the ARC TripleSpec (TSPEC) instrument. One additional object with an \xmm observation was selected from a publicly available, archival GNIRS 2004B dataset (project GS-2004B-Q16, PI Todd Boroson) that had not previously been reduced. This resulted in our final sample of 11 objects, listed in Table \ref{tab:sample}.

It is plausible that our selection criteria introduces some bias into the sample, for instance, by choosing objects with a higher than average IR/optical or X-ray/optical ratio. However, our mean value of $\alpha_{\rm OX}$ for Model 3 (see \S \ref{subsec:results}) is $1.61\pm0.09$, which is typical of large X-ray selected samples (\eg \citealt{vignali03}, \citealt{lusso10}). There may be an inherent bias in our sample owing to the flux threshold of the parent SDSS spectroscopic database. It is of course true that the sample must have X-ray data, which will naturally bias against some sub-classes of AGN, \eg BAL quasars.

\subsection{Optical/IR Data Preparation} \label{subsec:dataprep}
The IR spectra for the four objects selected from the \cite{shen12} sample were kindly provided by Yue Shen (private communication).

The data resulting from \gemini cycle 2013B were reduced according to the guidelines provided on the \gemini website, using \iraf \ {\sc v}2.14, and the \gemini \ \iraf package {\sc v}1.12. The archival GNIRS 2004B object (J0118$-$0052) was reduced before the release of {\sc v}1.12 and we thus used {\sc v}1.11.1 of the \gemini \ \iraf package to reduce this object.

All \gemini spectra were created by GNIRS in cross-dispersed (`XD') mode, and we were able to recover orders 3-8 for the 2013B objects. We recovered orders 3-6 in the 2004B object, due to a more limited range of flat fields provided. All stages of the reduction were visually inspected, to ensure no errors had occurred.

Telluric stellar spectra were provided by \gemini . The purpose of these spectra are to provide a well-defined reference spectrum that may be used to correct object spectra for non-constant sky transmission across the infrared wavelength range. To correct for telluric features, we were not able to use the {\sc xtellcor} routine discussed in \cite{vacca03} as not all of the telluric stars observed were of spectral class A0V. Hence we corrected for telluric features in the following way.

Hydrogen absorption features in the telluric spectrum were first removed in \iraf , as these features are intrinsic to the star itself. Lorentzian absorption line profiles were assumed, as these fit the features better than Gaussian or Voigt profiles. The assumed template for the telluric star was a blackbody of a characteristic temperature dependent on the telluric stellar class, and flux normalised to the 2MASS {\it J} magnitude. The extracted object spectrum was than multiplied by the ratio of the assumed telluric template to the extracted telluric spectrum, producing a very satisfactory correction to the variable atmospheric transmission.

Two factors may affect the relative normalisations of the IR and optical spectra, observed as they are at different epochs. Variability of the AGN is one possible origin for such a change, and will be discussed more in \S \ref{subsec:var}. The other factor is the accuracy of the flux calibration; if the optical or IR spectra are flux calibrated poorly (\eg due to seeing problems or aperture effects), the resulting normalisation of the spectra will be incorrect. For the SDSS/BOSS spectra, we will assume that the flux calibration is precise. 

The objects we reduced from \gemini were flux calibrated in the telluric correction step discussed above, which makes the assumption that the telluric star is not itself variable, and that weather conditions do not change between the object and the telluric observations. The Shen TSPEC objects, on the other hand, are all flux-normalised to the 2MASS {\it H} band magnitudes.

All of the objects in our sample have either UKIDSS or 2MASS (or both) photometry in each of {\it J}, {\it H}, and {\it K} bands. We first compare the flux of each photometry point with the average flux in the IR spectrum across the effective photometric bandpass, to estimate the percentage difference in flux density between these two values. As an additional check, we can also compare how well the blue wing of each IR spectrum fits a power-law extrapolated redwards of the SDSS spectrum, or, if available, the region of spectral overlap between optical/IR spectra. This highlights offsets of the IR spectrum relative to the optical spectrum. In doing this, we find that all but three of our objects show evidence for less than 10\% difference between the IR spectrum and optical spectrum/IR photometry.

The three objects showing greater than 10\% variation between optical and IR spectra are J0041$-$0947, J1044$+$2128 and J2328$+$1500.

In J1044$+$2128 and J2328$+$1500, we normalise to the {\it K}-band photometry, as there is evidence for some cloud cover at the time of observation that could affect the flux calibration. In J0041$-$0947, the IR spectrum was normalised to 2MASS {\it H} by Shen \& Liu. However, it lies below the level predicted from the SDSS spectrum, possibly due to variability. We thus make our mass estimate from the spectrum as flux-normalised by Shen \& Liu, but normalise the spectrum to the optical level for the SED fit, as whether the difference is due to flux calibration error in either spectrum, or variability, we require agreement to fit the SED shape.

Finally, in all objects we corrected both the IR and optical spectra for extinction by the Milky Way, using the dust maps of \cite{schlegel98} and the extinction law of \cite{cardelli89}.

All of our optical data come from the SDSS. Five of the objects had been observed multiple times in the SDSS spectroscopic survey, or reobserved by BOSS, the follow-up to SDSS-III. The BOSS survey offers greater wavelength coverage and improved S/N for the observed objects, and hence the BOSS spectra are our preferred source of optical data. In the absence of BOSS spectra, we use the SDSS spectrum that best aligns with the IR spectrum.

The optical and IR spectral observation dates are tabulated in Table \ref{tab:dates}. We provide estimates of the spectral S/N ratio by estimating and averaging the S/N in 10000 random, 50-pixel subsamples from each spectrum. The optical spectra adopted for the SED fitting are highlighted in bold. A greater discussion of variability in general is given in \S \ref{subsec:var}.

\subsection{UV and X-ray Data} \label{subsec:uv_xray}
We extracted all X-ray data from the \xmm Science Archive, through the HEASARC. For each object, we obtained spectra for all available observations from the EPIC MOS1, MOS2 and PN cameras, to maximise the number of counts. Each spectrum comes in three parts: a source spectrum, a background spectrum, and an ancillary response function. These are supplemented by an instrument-specific canned response matrix, downloaded from the \xmmn EPIC Response Files Page. These are all provided in a file format readable by \xspec .

For two of the objects in our sample for which IR spectra were pre-existing, the X-ray spectrum had not been extracted from the EPIC source image, due to lack of counts. However, the \xmmn Serendipitous Source Catalog (3XMM DR4) lists, for all detections, the X-ray flux in five bands: 0.2-0.5 keV, 0.5-1.0 keV, 1.0-2.0 keV, 2.0-4.5 keV, 4.5-12.0 keV, and we thus used these points in the SED fitting.

\xmmn also has an optical monitor (OM), which can obtain UV photometry simultaneously to the EPIC observation. Targets that are too far off-axis are not covered, so many of our sources that were observed serendipitously by EPIC do not have OM data.

Finally, we also searched the Galaxy Evolution Explorer (\galex) database for UV photometry. All sources have some data in the \galex All-sky Imaging Survey (AIS).

\section{Black Hole Masses} \label{sec:mass}
The first step in analysing data was to make BH mass estimates from the IR spectra. Discussion of a full spectral decomposition will be made in Paper II, but relies on the results of our broadband SED fits, so at this stage we will use localised decomposition of the Balmer line region, which is redshifted into our IR spectra.

\subsection{Optical/IR Spectral Fitting}

\begin{figure*}
\centering
\includegraphics[width = \textwidth]{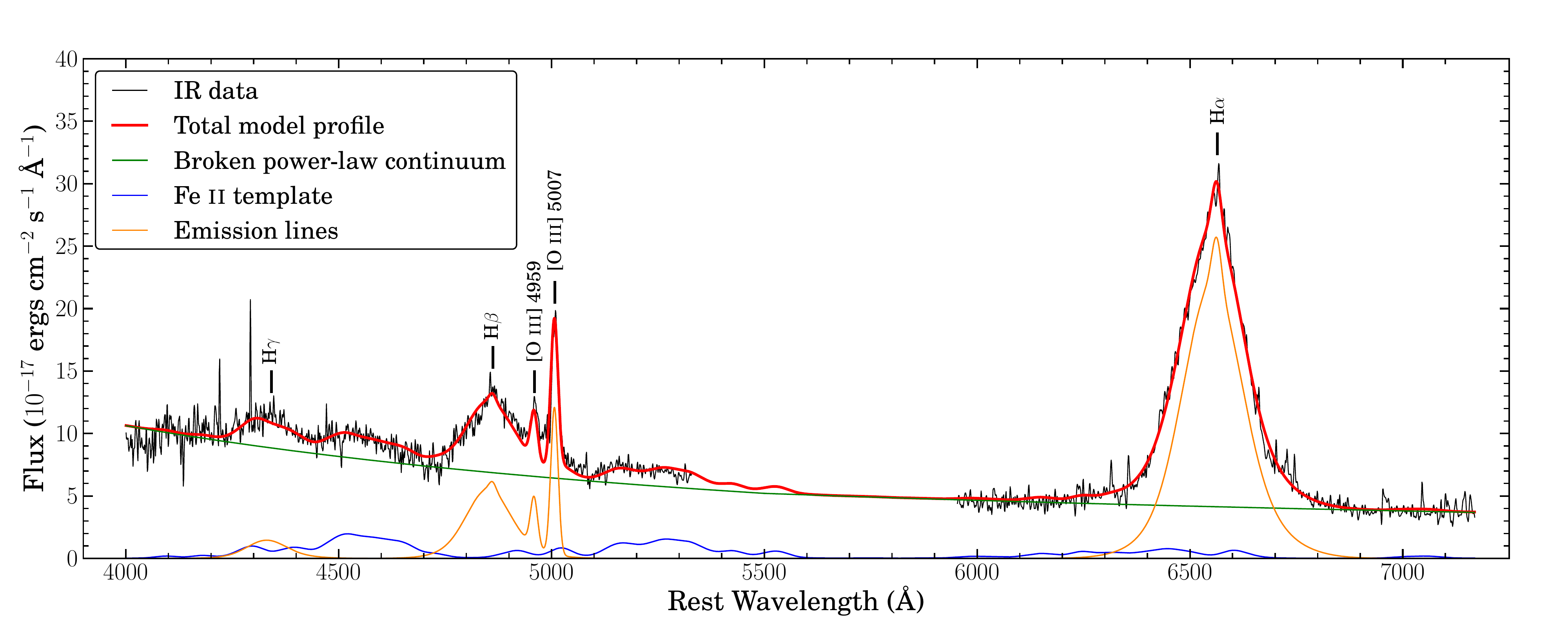}
\caption{\small Spectral decomposition of the Balmer lines for J2328$+$1500. This object shows reasonable S/N ($\sim$11) across the spectral range, whilst some of the other objects in our sample show significantly noisier spectra. Although we only use the FWHM and line luminosity of \Ha in making the BH mass estimate, we fit the other Balmer lines and the \Oiii doublet to more strongly constrain the fit, particularly of the iron emission. There is some evidence of a red wing to the \Hb profile. The extent of this and possible explanations will be explored in Paper II.}
\label{fig:balmer}
\end{figure*}

We first perform a continuum subtraction of the region under the Balmer lines. Whilst the underlying, featureless continuum is in theory best described by an accretion disc, on localised scales it may approximate a power-law continuum. As discussed in \eg \cite{vandenberk01} the continuum under the Balmer lines is more accurately represented by a broken power law, so in our Balmer region continuum fit we also employ a broken power-law. We use, for the power-law, the standard form:
\begin{equation} \label{eq1}
F(\lambda) = C_1(\lambda / 5100 \, \textrm{\AA})^{-C_2}
\end{equation}
where $C_1$ is the normalisation and $C_2$ is the power-law slope, and in our case, implement the change in index at 5500 \AA \ rest-frame.

Our objects are all relatively bright, and so contamination from the host galaxy is expected to be negligible (see \S \ref{subsec:host}), however AGN spectra are also contaminated across the spectral range by broad, blended \Feii multiplets. Modelling of these features is often accomplished by making use of an empirical template derived from spectral analysis of the Type I AGN I Zwicky 1. We use that of \cite{veroncetty04}. There are two free parameters; the width of the convolving Gaussian, and the normalisation. We fit the broken power-law and \Feii pseudo-continuum to the spectral regions in between the strong emission lines -- we use: 4000$-$4300, 4400$-$4750, 5050$-$5800, 5950$-$6300 and 6800$-$7500 \AA . Our fitted continuum is subtracted to leave the Balmer emission line spectrum.

We then fitted the broad permitted emission lines \Ha , \Hb and \Hg and the narrow, forbidden lines \Oiii . We see two categories of object in our sample; three show strong, narrow \Oiii lines, and the other eight do not. This will be discussed more in Paper II. Following common practice, we fit multiple Gaussian components to emission lines, which offers a reasonable approximation to the line shape, and given the quality of our data, is perfectly adequate for our purpose. Lines are fitted as follows:

\begin{itemize}
\item[i.]{\Ha is fitted with two Gaussian components -- one broad and one intermediate. These are free in velocity shift and normalisation. For the three objects that show strong narrow \Oiii , a third, narrow Gaussian component, of wavelength and velocity width tied to the strong member of the \Oiii doublet is also included.}
\item[ii.]{\Hb is fitted with two Gaussians, with wavelengths and velocity widths tied to the corresponding components in \Ha and with the same amplitude ratio. A narrow component is included if strong \Oiii is observed.}
\item[iii.]{\Hg is fitted for completeness with 2 components, tied in amplitude ratio, velocity width and wavelength to the corresponding components in \Ha .}
\item[iv.]{\Oiii is a doublet; each member is fitted with a single Gaussian, tied together in velocity width, and with an amplitude ratio of 2.98 \citep{storey00}.}
\end{itemize}

We obtain the BH mass estimate for our objects from the \Ha line, using the method derived by \cite{greene05}. It is common to use \Hb , as much study has been devoted to calibrating mass estimates from \Hb with RM samples, due to the greater availability of \Hb in optical spectra (\Ha being redshifted to IR wavelengths in AGN above $z \sim 0.3$), and also because \Ha profiles may be blended with narrow forbidden lines \Sii and \Nii . However, given the limited S/N of our data, we prefer to use the $\sim 3 \times$ stronger \Ha profile. We note that we do not detect the \Sii doublet in any of our objects (even those showing \Oiii ), which is supporting evidence that contamination of the \Ha profile by \Nii is likely to be small. Whilst \Hb mass estimates are made using FWHM$_{\rm H \beta}$ as a proxy for the velocity dispersion, and the continuum luminosity at 5100 \AA \ ($L_{5100 \rm \mathring{A}}$) as a proxy for the BLR size, \cite{greene05} found strong correlations between FWHM$_{\rm H \beta}$ and FWHM$_{\rm H \alpha}$ and between $L_{5100 \rm \mathring{A}}$ and $L_{\rm H \alpha}$, and so used this as the basis for derivation of their relation between \Ha profile and BH mass:

\begin{equation} \label{eq2}
\begin{split}
M_{\rm BH} & = (2.0^{+0.4}_{-0.3}) \times 10^6 \\
 & \times \left( \frac{L_{\rm H \alpha}}{10^{42} \, \rm erg \, s^{-1}} \right)^{0.55 \pm 0.02} \! \!
\left( \frac{\rm FWHM_{H \alpha}}{10^3 \, \rm km \, s^{-1}} \right)^{2.06 \pm 0.06} \! \! M_{\odot} .
\end{split}
\end{equation}

Following the method of \cite{greene05} and others, we use the full, broad \Ha component to obtain BH mass estimates. J12a discussed the merits of drawing a distinction between the Gaussian components that form the broad profile, defining these as `broad' and `intermediate' components, but eventually they determined that the two combined yielded the most reliable BH mass estimates.

\subsection{Results}
We use a Levenberg-Marquardt minimisation method throughout. An example of the Balmer region spectral decomposition is shown in Fig.\ \ref{fig:balmer}. Measurement errors are estimated using a Monte Carlo technique, where different iterations of the data are generated using the `mean' (measured) flux value and the error on that value. For each different iteration, optimal values are recalculated, and this is repeated 100 times to estimate the error of each fitted value. This procedure is not perfect, as we add noise to the already noise-degraded spectral data, but serves as a suitable approximation. The fitting errors are not indicative of the error on the resulting mass estimate though, which are made in accordance with Equation \ref{eq2}. We tabulate the results of the emission line analysis in Table \ref{tab:line}.

\begin{table}
\caption{\small \Ha line measurements for the sample. For objects common to the Shen and Liu (2012) sample, our measurements agree to within 2$\sigma$.}
\small
\centering
\begin{tabular}{cccccc}
\hline
ID & FWHM$_{\rm H\alpha}$ & $\log$ & $z_{\rm [O{\sc III}]}$ & $\log$\\
    & (km s$^{-1}$)        &   $(L_{\rm H\alpha }/$erg s$^{-1})$ &  & $(M_{\rm BH}/M_{\odot})$\\
\hline
1   & 5600$\pm$200  & 44.85$\pm$0.02 &  1.629 & 9.42$\pm$0.10   \\
2   & 2940$\pm$90  & 44.57$\pm$0.01 &   1.567 & 8.68$\pm$0.09   \\
3   & 4680$\pm$60  & 44.86$\pm$0.01 &   2.192 & 9.25$\pm$0.10   \\
4   & 3100$\pm$100  & 44.40$\pm$0.04 &  1.545 & 8.63$\pm$0.09   \\
5   & 5100$\pm$200 & 45.21$\pm$0.01 &   1.535 & 9.53$\pm$0.10   \\
6   & 3200$\pm$100  & 44.53$\pm$0.01 &  1.577 & 8.73$\pm$0.09   \\
7   & 2550$\pm$80  & 44.56$\pm$0.01 &   1.500 & 8.55$\pm$0.09   \\
8   & 2460$\pm$20  & 44.87$\pm$0.01 &   1.562 & 8.68$\pm$0.09   \\
9   & 3390$\pm$80  & 44.94$\pm$0.02 &   1.623 & 9.01$\pm$0.09   \\
10  & 7810$\pm$80  & 44.81$\pm$0.01 &   1.539 & 9.68$\pm$0.10   \\
11  & 5000$\pm$70  & 44.32$\pm$0.01 &   1.609 & 9.02$\pm$0.09   \\
\hline
\end{tabular}
\label{tab:line}
\end{table}

\section{Bolometric Luminosity}
With the BH mass estimates in hand, we could make predictions as to the bolometric luminosity of each AGN by fitting the broadband SED model \optxagnf to the optical/IR and X-ray data for each object. We will take the bolometric luminosity to mean the total intrinsic luminosity of the nuclear source, excluding stars in the host galaxy and any associated reradiation.

\subsection{X-ray Spectrum} \label{subsec:xrayfit}
We first fit the X-ray data with an absorbed power-law. We include attenuation attributable to both the Milky Way (fixed) and the host galaxy (free). This will allow us to verify that the following section, in which we fit the \cite{done12} \optxagnf model, gives reasonable values. We calculate Milky Way \nh \ values using the Leiden/Argentine/Bonn Survey of Galactic HI (\citealt{kalberla05}).

In Table \ref{tab:xray} the X-ray exposure times and count values are tabulated, and the fitted parameters are shown together with 90\% confidence limits. 

\begin{table*}
\caption{\small The model properties for the X-ray spectrum. We fit absorption components for both the Milky Way (fixed) and the host galaxy (free). In many objects the host absorption is poorly constrained, due to the high redshift (only the tail of the absortion profile is sampled) and limited number of counts. The EPIC count errors are 1 $\sigma$, and the errors on the model parameters are the 90\% confidence limits, in line with convention in X-ray astronomy. For similar reasons, we also quote $\Gamma$ and its uncertainty to two decimal places.}
\small
\centering
\begin{tabular}{ccccccc}
\hline
ID & Exp. Time           & \xmm EPIC Cts    & $N_{\rm H, \, MW}$  & $N_{\rm H,\,int}$
   & $\Gamma$            & \rchi    \\
   & (s)           &           & ($\times 10^{20}\,$cm$^{-3}$)  & ($\times 10^{20}\,$cm$^{-3}$)
   &                     &    \\
\hline
1  & 12 519              & 410$\pm$20       & 2.64                & 0$^{+16}_{-0}$	  
   & 2.34$^{+0.24}_{-0.22}$& 1.29  \\
2  & 21 304              & 850$\pm$30       & 1.83                & 0$^{+11}_{-0}$	  
   & 2.56$^{+0.19}_{-0.16}$& 0.89 \\
3  & 10 523              & 142$\pm$17       & 3.89                & 50$^{+110}_{-50}$   
   & 2.92$^{+0.79}_{-0.52}$& 1.71  \\
4  & 7 179               & 203$\pm$18       & 2.58                & 11$^{+18}_{-11}$    
   & 2.07$^{+0.12}_{-0.13}$& 0.60 \\
5  & 15 781              & 7200$\pm$100     & 4.48                & 45$^{+13}_{-12}$    
   & 1.99$^{+0.09}_{-0.08}$& 1.32 \\
6  & 19 591              & 550$\pm$30       & 4.04                & 20$^{+40}_{-20}$    
   & 2.33$^{+0.41}_{-0.31}$& 0.77  \\
7  & 154 071             & 4950$\pm$80      & 1.73                & 0$^{+5}_{-0}$	      
   & 2.30$^{+0.07}_{-0.06}$& 1.54 \\
8  & 8 117               & 760$\pm$30       & 1.31                & 0$^{+19}_{-0}$	  
   & 1.80$^{+0.17}_{-0.11}$& 1.45 \\
9  & 23 543              & 1130$\pm$40      & 1.24                & 0$^{+17}_{-0}$	  
   & 2.20$^{+0.19}_{-0.12}$& 0.84 \\
10 & 124 956             & 1350$\pm$50      & 3.85                & 0$^{+13}_{-0}$	  
   & 1.44$^{+0.12}_{-0.11}$& 1.14 \\
11 & 34 705              & 710$\pm$30       & 4.00                & 0$^{+19}_{-0}$      
   & 2.19$^{+0.12}_{-0.13}$& 0.62 \\
\hline

\end{tabular}
\label{tab:xray}
\end{table*}

\subsection{Broadband SED Fitting} \label{subsec:sedfit}
There are a number of properties that affect the intrinsic AGN SED; these are described in \cite{done12}. The observed SED is also affected by optical/UV extinction and soft X-ray absorption due to interstellar dust and photoelectric absorption (respectively) in the Milky Way. We correct for the extinction (reddening) in the Milky Way as discussed in \S \ref{subsec:dataprep}, and as in \S \ref{subsec:xrayfit} include a photoelectric absorption component (\wabs) to model the soft X-ray absorption by Hydrogen, Helium, etc.

The AGN host galaxy is assumed to have similar intrinsic processes reddening/absorbing its emission. We can also model these components by redshifting models for X-ray absorption (\zwabs) and extinction (\zredden), albeit with no means of constraining these other than the shape of the SED. We therefore also produce a second model in which these components are added.

In this redshift range our data may not be sufficient to constrain the properties of the SX. Equally, it is possible that either our X-ray data or optical data may sample this part of the SED, depending on the coronal radius. Since we know empirically (\eg J12a) that the SX ought to be taking up a significant fraction ($\sim 70 \%$) of the Comptonised energy, we fix its properties to reasonable default values, detailed below.

In both scenarios, the constrained parameters are as follows. A discussion of specific exceptions follows.

\begin{itemize}
\item[i.]{BH Mass, \M_BH : as previously described in \S \ref{sec:mass}, we constrain BH mass using the method of \cite{greene05} based on \Ha .}
\item[ii.]{Redshift, $z$: as measured in the spectral decomposition.}
\item[iii.]{Distance, $r_{\rm c}$: we calculate the comoving distance to each source from the measured redshift assuming a flat cosmology (with ${H_0=70 \, \rm km \, s^{-1} Mpc^{-1}}$, ${\Omega_{\rm M}=0.27}$ and ${\Omega_{\Lambda}=0.73}$).}
\item[iv.]{BH spin, $a_\star$: initially we will constrain spin to be zero, in line with the work of J12a, but specific instances where a spinning BH is possible or implied will be explored in Paper II.}
\item[v.]{Electron temperature of SX, $kT_{\rm e}$: fixed at a typical value of 0.2 keV.}
\item[vi.]{Optical depth of SX, $\tau$: fixed at a typical value of 10.}
\item[vii.]{Fraction of Comptonised component in SX, $f_{\rm SX}$: fixed at a typical value of 0.7.}
\item[viii.]{Hydrogen column density (Milky Way), $N_{\rm H, \, gal}$: calculated using the Leiden/Argentine/Bonn Survey of Galactic HI (\citealt{kalberla05})}
\end{itemize}

The fitted parameters are as follows.
\begin{itemize}
\item[i.]{Mass accretion rate, $\dot m = L_{\rm bol}/L_{\rm Edd}$.}
\item[ii.]{Coronal radius, $r_{\rm cor}$: the radius at which the AD energy is reprocessed by SX and PLT.}
\item[iii.]{Radial extent of AD, $r_{\rm out}$: the outer radius, in $R_{\rm g}$ of the AD. In some objects this cannot be constrained -- in these $r_{\rm out}$ is given as a limit.}
\item[iv.]{Power-law slope, $\Gamma$: The power-law index of the coronal PLT.}
\item[v.]{Intrinsic Hydrogen column density, $N_{\rm H, \, int}$ (\zwabs): soft X-ray attenuation intrinsic to the host galaxy.}
\item[vi.]{Intrinsic reddening, E$(B-V)$ (\zredden): redshifted extinction curve to account for reddening intrinsic to the host galaxy.}
\end{itemize}

We also produce for each object a final model in which the SX normalisation is permitted to vary. This allows us to test the hypothesis that this component is constrained in some objects.

To summarise, the models fitted are:

\begin{itemize}
\item[i.]{Model 1: SX fixed, no intrinsic attenuation}
\item[ii.]{Model 2: SX fixed, incl.\ intrinsic attenuation}
\item[iii.]{Model 3: SX free, incl.\ intrinsic attenuation}
\end{itemize}

Defining the data to be fit in \xspec is not straightforward. The optical/IR spectral data is, as previously mentioned, contaminated by emission features, including complex, blended \Feii emission, and the Balmer continuum. We therefore selected, using the \cite{vandenberk01} quasar template, spectral regions free from such emission, and binned these narrow wavelength ranges up into well-defined photometry points. The error on each point is defined as the flux density standard deviation across the bin. The ranges used are (where available): 1300$-$1350, 1425$-$1475, 1700$-$1750, 2175$-$2225, 3900$-$4000, 4150$-$4250, 5600$-$5700, 6100$-$6200, 6900$-$7000, unless any of these were unsuitable, \eg for reasons of poor S/N. We then fitted the full energy range of available X-ray data from \xmm EPIC.

The bluest bins chosen also do not cover the absorption features observed by \cite{kaastra14}, which may be expected in AGN with high $N_{\rm H}$ columns (see Table \ref{tab:xray}). We do not tie the $N_{\rm H, int}$ and intrinsic E$(B-V)$ values together.

The UV photometry from \xmm OM and \galex was not included in the modelling process. At the redshifts we are considering, the UV filters on these observatories cover a broad wavelength range over the strong Lyman-$\alpha$ $\lambda$1216 \AA \ emission line, and the Ly-$\alpha$ forest beyond, and are therefore not a good indication of the continuum level. With high-resolution UV spectral data, it would be possible to interpolate over the narrow absorption features in the forest, and recover the underlying continuum (\eg \citealt{finn14}) but this would require observations with \eg {\it HST}/COS.

\subsection{Results} \label{subsec:results}
The results of the SED fitting procedure are as follows. In Table \ref{tab:sedprop} the best-fitting parameters for each model are tabulated, including the mass accretion rate in $M_{\odot}$ yr$^{-1}$. In Table \ref{tab:modprop} the key properties of these SED models are listed. In the manner of J12, we have calculated $\kappa_{\rm 2-10\,keV} = L_{\rm bol} / L_{\rm 2-10\,keV}$ and $\kappa_{5100 \rm \mathring{A}} = L_{\rm bol} / L_{5100 \rm \mathring{A}}$, the $2-10$ keV and $5100$ \AA \ bolometric correction coefficients, as these are commonly used proxies for the bolometric luminosity. We also give $\alpha_{\rm OX}$ (\eg \citealt{lusso10}). The uncertainties quoted are the 90\% confidence limits, as is conventional in X-ray astronomy. We have estimated these using the Fisher matrix, which gives an indication of the measurement error. It should be remembered that this does not take account of the systematic errors, to which the main contributors will be the uncertainties on the mass estimate and the flux calibration.

We note that some of the models show the SX component to be unconstrained by the data, and in five objects this is manifested by large fitting errors on \rcor and \fsx in Model 3. For Model 3, in the interest of limiting the number of free parameters, and allowing the model to converge to a meaningful minimum, we lock $N_{\rm H, \, int}$, and $r_{\rm out}$ to the Model 2 values. We make an exception to this rule for J0839$+$5754, as the X-ray data is sufficient to well-constrain $N_{\rm H, \, int}$.

\begin{table*}
\caption{\small The optimum fitted parameters for the various SED models. Uncertainties quoted are the 90\% confidence limits, as is conventional in X-ray astronomy, and are estimated using the Fisher matrix. As such, they are only indicative of the true measurement error.}
\small
\centering
\begin{tabular}{cccccccccc}
\hline
ID & $N_{\rm H, \, int}$ & E$(B-V)$         & $\dot{m}=$       & $\dot{M}$     & $r_{\rm cor}$ 
   & $r_{\rm out}$       & $\Gamma$         & $f_{\rm SX}$     & $\chi^2_{\rm reduced}$ \\
   & ($10^{22} \,$cm$^{-2}$) & (mag)        & \Lbol$/$\LEdd    & ($M_{\odot}$ yr$^{-1}$)
   & ($R_{\rm g}$)& ($R_{\rm g}$)&          &                  \\
\hline
\multicolumn{9}{l}{\textbf{Model 1: No intrinsic attenuation, SX fixed} (SX parameters: $kT_{\rm e}=0.2$ keV, $\tau=10$, $f_{\rm SX}=0.7$)} \\
\hline
1  & (0.0)               & (0.0)            & 0.389$\pm$0.018  & 44$\pm$2      & 23$\pm$3    
   & $>$1000             & 2.21$\pm$0.03    & (0.7)            & 10.4   \\
							       			     
2  & (0.0)               & (0.0)            & 2.79$\pm$0.05    & 56.8$\pm$1.0  & 11.6$\pm$0.6
   & 790$\pm$80          & 2.53$\pm$0.09    & (0.7)            & 0.97   \\
   							       			     
3  & (0.0)               & (0.0)            & 0.39$\pm$0.06    & 30$\pm$5      & 24$\pm$9
   & 300$\pm$60          & 2.45$\pm$0.03    & (0.7)            & 2.39   \\
							       			     
4  & (0.0)               & (0.0)            & 2.22 $\pm$0.07   & 40.3$\pm$1.4  & 9.87$\pm$0.16
   & 330$\pm$30          & 1.94$\pm$0.05    & (0.7)            & 3.61   \\
							       			        
5  & (0.0)               & (0.0)            & 0.231$\pm$0.018  & 33$\pm$3      & 88.3$\pm$1.4 
   & $>1000$             & 1.86$\pm$0.04    & (0.7)            & 3.50   \\
							       			     
6  & (0.0)               & (0.0)            & 1.24$\pm$0.03    & 28.3$\pm$0.6  & 13.2$\pm$0.6 
   & 1800$\pm$600        & 2.14$\pm$0.11    & (0.7)            & 2.60   \\
							       			     
7  & (0.0)               & (0.0)            & 2.9$\pm$0.2      & 44$\pm$3      & 10.2$\pm$0.4
   & $>$10000            & 2.27$\pm$0.03    & (0.7)            & 2.63   \\

8  & (0.0)               & (0.0)            & 2.10$\pm$0.04    & 42.7$\pm$0.8  & 15.9$\pm$1.1 
   & $>$10000            & 1.79$\pm$0.07    & (0.7)            & 6.38   \\
							       			     
9  & (0.0)               & (0.0)            & 1.410$\pm$0.016  & 60.1$\pm$0.7  & 10.8$\pm$0.3
   & 700$\pm$60          & 2.18$\pm$0.07    & (0.7)            & 1.78   \\
							       			     
10 & (0.0)               & (0.0)            & 0.0371$\pm$0.0011& 7.6$\pm$0.2   & 21$\pm$3 
   & 56.0$\pm$1.9        & 1.63$\pm$0.06    & (0.7)            & 1.77   \\
							       			     
11 & (0.0)               & (0.0)            & 0.317$\pm$0.010  & 14.2$\pm$0.5  & 20$\pm$3
   & 237$\pm$14          & 2.18$\pm$0.08    & (0.7)            & 0.89   \\
 
\hline
\multicolumn{9}{l}{\textbf{Model 2: Incl.\ intrinsic attenuation, SX fixed} (SX parameters: $kT_{\rm e}=0.2$ keV, $\tau=10$, $f_{\rm SX}=0.7$)}\\
\hline
1  & 0.0$\pm$0.3         & 0.051$\pm$0.006  & 0.61$\pm$0.12    & 68$\pm$14     & 25$\pm$9
   & 240$\pm$40          & 2.38$\pm$0.03    & (0.7)            & 5.01   \\
   							       	       
2  & 0.0$\pm$0.3         & 0.015$\pm$0.010  & 3.3$\pm$0.4      & 67$\pm$8      & 10.9$\pm$1.3 
   & 790$\pm$60          & 2.52$\pm$0.19    & (0.7)            & 0.95   \\
							       	       
3  & 0.15$\pm$0.12       & 0.025$\pm$0.015  & 0.50$\pm$0.14    & 38$\pm$11     & 25$\pm$11 
   & 260$\pm$60          & 2.50$\pm$0.06    & (0.7)            & 2.48   \\
							       	       
4  & 0.20$\pm$0.10       & 0.060$\pm$0.015  & 4.1$\pm$0.7      & 75$\pm$12     & 9.0$\pm$0.3 
   & 400$\pm$40          & 2.07$\pm$0.07    & (0.7)            & 0.51   \\
							       	       
5  & 0.18$\pm$0.06       & 0.065$\pm$0.003  & 0.338$\pm$0.007  & 48.8$\pm$1.0  & 80.1$\pm$1.2 
   & $>$1000             & 1.86$\pm$0.05    & (0.7)            & 1.45   \\
							       	       
6  & 0.2$\pm$0.2         & 0.024$\pm$0.010  & 1.58$\pm$0.16    & 36$\pm$4      & 13$\pm$3  
   & 970$\pm$130         & 2.32$\pm$0.22    & (0.7)            & 2.52   \\
							       	       
7  & 0.00$\pm$0.11       & 0.033$\pm$0.006  & 4.1$\pm$0.3      & 62$\pm$5       & 9.4$\pm$0.2 
   & $>$10000            & 2.26$\pm$0.06    & (0.7)            & 2.54   \\
							       	       
8  & 0.01$\pm$0.17       & 0.052$\pm$0.014  & 3.06$\pm$0.12    & 62$\pm$2      & 13$\pm$2 
   & $>$10000            & 1.80$\pm$0.13    & (0.7)            & 1.77   \\
							       	       
9  & 0.0$\pm$0.3         & 0.030$\pm$0.008  & 1.98$\pm$0.18    & 84$\pm$7      & 9.8$\pm$0.5 
   & 500$\pm$30          & 2.18$\pm$0.13    & (0.7)            & 1.62   \\
							       	       
10 & 0.0$\pm$0.3         & 0.094$\pm$0.015  & 0.067$\pm$0.007  & 13.6$\pm$1.4  & 19$\pm$4 
   & 53$\pm$2            & 1.50$\pm$0.09    & (0.7)            & 1.30   \\
							       	       
11 & 0.0$\pm$0.3         & 0.023$\pm$0.015  & 0.42$\pm$0.08    & 19$\pm$3      & 16$\pm$4 
   & 211$\pm$17          & 2.18$\pm$0.14    & (0.7)            & 0.88   \\

\hline
\multicolumn{9}{l}{\textbf{Model 3: Incl.\ intrinsic attenuation, SX free} (SX parameters: $kT_{\rm e}=0.2$ keV, $\tau=10$, $f_{\rm SX}=\,$free)}\\
\hline
1  & (0.0)               & 0.051$\pm$0.005  & 0.60$\pm$0.04    & 67$\pm$4      & 25$\pm$2   
   & (240)               & 2.17$\pm$0.21    & 0.83$\pm$0.09    & 4.69   \\
   							       	       
2  & (0.0)               & 0.011$\pm$0.008  & 3.2$\pm$0.3      & 64$\pm$6      & 15$\pm$40* 
   & (790)               & 2.48$\pm$0.14    & 0.9$\pm$0.3*     & 0.93   \\
							       	       
3  & (0.15)              & 0.028$\pm$0.013  & 0.51$\pm$0.05    & 38$\pm$4      & 26$\pm$5  
   & (260)               & 2.47$\pm$0.09    & 0.74$\pm$0.10    & 2.14   \\
					       	       
4  & (0.20)              & 0.060$\pm$0.014  & 4.1$\pm$0.7      & 75$\pm$12     & 9.0$\pm$1.1 
   & (400)               & 2.07$\pm$0.06    & 0.7$\pm$0.2      & 0.42   \\
								       	       
5  & 0.51$\pm$0.07       & 0.051$\pm$0.006  & 0.336$\pm$0.009  & 48.6$\pm$1.3  & 82.7$\pm$1.7 
   & ($>$1000)           & 2.06$\pm$0.05    & 0.56$\pm$0.05    & 1.29   \\
							       	       
6  & (0.2)               & 0.024$\pm$0.008  & 1.58$\pm$0.07    & 36.1$\pm$1.5  & 10$\pm$80* 
   & (970)               & 2.32$\pm$0.14    & 0.7$\pm$3*       & 2.38   \\
							       	       
7  & (0.0)               & 0.033$\pm$0.006  & 4.1$\pm$0.3      & 62$\pm$4      & 9.9$\pm$1.8 
   & ($>$10000)          & 2.25$\pm$0.05    & 0.8$\pm$0.2      & 2.53   \\
							       	       
8  & (0.01)              & 0.056$\pm$0.004  & 3.16$\pm$0.10    & 64$\pm$2      & 10$\pm$10* 
   & ($>$10000)          & 1.80$\pm$0.08    & 0.7$\pm$0.7*     & 1.70   \\
							       	       
9  & (0.0)               & 0.030$\pm$0.005  & 1.97$\pm$0.11    & 84$\pm$5      & 10$\pm$14*   
   & (500)               & 2.19$\pm$0.09    & 0.7$\pm$2*       & 1.58   \\
							       	       
10 & (0.0)               & 0.091$\pm$0.015  & 0.066$\pm$0.006  & 13.5$\pm$1.3  & 17$\pm$4  
   & (53)                & 1.49$\pm$0.06    & 0.64$\pm$0.16    & 1.28   \\
						       	       
11 & (0.0)               & 0.023$\pm$0.009  & 0.42$\pm$0.05    & 19$\pm$2      & 15$\pm$40*  
   & (211)               & 2.18$\pm$0.08    & 0.7$\pm$1.2*     & 0.85   \\

\hline
\multicolumn{9}{l}{*Large error indicative of unconstrained SX parameter}

\end{tabular}
\label{tab:sedprop}
\end{table*}

\begin{table*}
\caption{\small The key properties of the various SED models, including bolometric correction coefficients.}
\small
\centering
\begin{tabular}{cccccccccc}
\hline
ID & $\log (L_{\rm bol})$       & $\log (L_{\rm 2-10 \, keV})$    & $\kappa_{\rm 2-10\,keV}$ 
   & $\log(\lambda L_{2500 \rm \mathring{A}})$& $\log(\nu L_{2\,\rm keV})$ & $\alpha_{\rm OX}$  
   & $\log(\lambda L_{5100 \rm \mathring{A}})$& $\kappa_{5100 \rm \mathring{A}}$ \\
   
   &       \multicolumn{2}{c}{$(\log$(erg s$^{-1}$)$)$}           &  
   &       \multicolumn{2}{c}{$(\log$(erg s$^{-1}$)$)$}           &  
   & $(\log$(erg s$^{-1}$)$)$   &                                  \\
\hline
\multicolumn{9}{l}{\textbf{Model 1: No intrinsic attenuation, SX fixed} (SX parameters: $kT_{\rm e}=0.2$ keV, $\tau=10$, $f_{\rm SX}=0.7$)}  \\
\hline
1  & 47.17$\pm$0.02             & 45.28                           & 76.7 
   & 46.70                      & 45.15                           & 1.60
   & 46.43                      & 5.42       \\
2  & 47.271$\pm$0.007           & 44.86                           & 257  
   & 46.36                      & 44.83                           & 1.59
   & 45.99                      & 18.9       \\
3  & 46.98$\pm$0.07             & 44.78                           & 159  
   & 46.49                      & 44.72                           & 1.68
   & 46.16                      & 6.56       \\
4  & 47.103$\pm$0.015           & 44.78                           & 212  
   & 46.15                      & 44.55                           & 1.62
   & 45.62                      & 30.3       \\
5  & 47.07$\pm$0.03             & 45.69                           & 24.0 
   & 46.68                      & 45.43                           & 1.48
   & 46.56                      & 3.20       \\
6  & 46.977$\pm$0.009           & 44.94                           & 109  
   & 46.18                      & 44.78                           & 1.54
   & 45.89                      & 12.1       \\
7  & 47.17$\pm$0.03             & 44.78                           & 242  
   & 46.20                      & 44.67                           & 1.59
   & 45.91                      & 18.0       \\
8  & 47.163$\pm$0.008           & 45.31                           & 71.8 
   & 46.27                      & 45.02                           & 1.48
   & 45.99                      & 15.0       \\
9  & 47.293$\pm$0.005           & 44.99                           & 202  
   & 46.57                      & 44.85                           & 1.66
   & 46.26                      & 10.9       \\
10 & 46.255$\pm$0.013           & 44.60                           & 45.3 
   & 45.97                      & 44.25                           & 1.66
   & 45.73                      & 3.38       \\
11 & 46.646$\pm$0.014           & 44.79                           & 72.4 
   & 46.11                      & 44.65                           & 1.56
   & 45.73                      & 8.16       \\

\hline
\multicolumn{9}{l}{\textbf{Model 2: Incl.\ intrinsic attenuation, SX fixed} (SX parameters: $kT_{\rm e}=0.2$ keV, $\tau=10$, $f_{\rm SX}=0.7$)}\\
\hline
1  & 47.33$\pm$0.09             & 45.27                           & 116  
   & 46.85                      & 45.19                           & 1.63
   & 46.48                      & 7.15    \\
2  & 47.36$\pm$0.05             & 44.86                           & 312  
   & 46.41                      & 44.83                           & 1.61
   & 46.02                      & 21.7    \\
3  & 47.08$\pm$0.12             & 44.82                           & 181  
   & 46.56                      & 44.78                           & 1.69
   & 46.20                      & 7.56    \\
4  & 47.38$\pm$0.07             & 44.86                           & 332  
   & 46.34                      & 44.68                           & 1.64
   & 45.80                      & 37.7    \\
5  & 47.233$\pm$0.009           & 45.84                           & 24.6 
   & 46.86                      & 45.58                           & 1.49
   & 46.65                      & 3.80    \\
6  & 47.08$\pm$0.05             & 44.95                           & 135  
   & 46.25                      & 44.85                           & 1.54
   & 45.94                      & 13.7    \\
7  & 47.32$\pm$0.03             & 44.78                           & 348  
   & 46.31                      & 44.66                           & 1.63
   & 46.02                      & 20.1    \\
8  & 47.323$\pm$0.017           & 45.30                           & 106  
   & 46.39                      & 45.01                           & 1.53
   & 46.10                      & 16.7    \\
9  & 47.43$\pm$0.04             & 44.99                           & 278  
   & 46.67                      & 44.85                           & 1.70
   & 46.29                      & 13.8    \\
10 & 46.50$\pm$0.05             & 44.63                           & 73.0 
   & 46.22                      & 44.23                           & 1.76
   & 45.86                      & 4.32    \\
11 & 46.76$\pm$0.08             & 44.79                           & 93.5 
   & 46.19                      & 44.65                           & 1.59
   & 45.77                      & 9.77    \\

\hline
\multicolumn{9}{l}{\textbf{Model 3: Incl.\ intrinsic attenuation, SX free} (SX parameters: $kT_{\rm e}=0.2$ keV, $\tau=10$, $f_{\rm SX}=\,$free)}\\
\hline
1  & 47.33$\pm$0.03             & 45.30                           & 106  
   & 46.84                      & 45.16                           & 1.65
   & 46.48                      & 7.12    \\
2  & 47.33$\pm$0.04             & 44.87                           & 290  
   & 46.39                      & 44.82                           & 1.60
   & 46.02                      & 20.2    \\
3  & 47.08$\pm$0.04             & 44.82                           & 185  
   & 46.57                      & 44.77                           & 1.69
   & 46.20                      & 7.62    \\
4  & 47.38$\pm$0.07             & 44.86                           & 332  
   & 46.34                      & 44.68                           & 1.64
   & 45.80                      & 37.7    \\
5  & 47.232$\pm$0.011           & 45.89                           & 22.2 
   & 46.81                      & 45.70                           & 1.43
   & 46.63                      & 4.01    \\
6  & 47.078$\pm$0.018           & 44.95                           & 136  
   & 46.25                      & 44.85                           & 1.54
   & 45.94                      & 13.7    \\
7  & 47.32$\pm$0.03             & 44.78                           & 348  
   & 46.31                      & 44.66                           & 1.63
   & 46.02                      & 20.1    \\
8  & 47.337$\pm$0.014           & 45.29                           & 110  
   & 46.40                      & 45.01                           & 1.53
   & 46.11                      & 16.9    \\
9  & 47.43$\pm$0.03             & 44.99                           & 278  
   & 46.67                      & 44.85                           & 1.70
   & 46.29                      & 13.8    \\
10 & 46.49$\pm$0.04             & 44.64                           & 71.4 
   & 46.21                      & 44.23                           & 1.76
   & 45.86                      & 4.28    \\
11 & 46.76$\pm$0.05             & 44.79                           & 93.4 
   & 46.19                      & 44.65                           & 1.59
   & 45.77                      & 9.76    \\

\hline

\end{tabular}
\label{tab:modprop}
\end{table*}

All the SED models are plotted, together with the observational data, in Appendices \ref{app:sed1}, \ref{app:sed2} and \ref{app:sed3}.

We see a rather limited range of SED shapes, with all but one object being disc-dominated, similar to the SEDs of NLS1s in the J12 sample. The lowest mass objects, J0043$+$0114, J0157$-$0048, J1021$+$1315, J1044$+$2128, J1240$+$4740 and J1350$+$2652 have unsampled SED peaks. The red wing of the accretion disc is better constrained than for the J12 objects though, owing to the lower fraction of host galaxy contribution in these high luminosity quasars.

The objects with BH masses $\gtrsim \! \! 10^9 M_{\odot}$ -- J0041$-$0947, J0118$-$0052, J0839$+$5754, J2328$+$1500 and J2332$+$0000 -- all have observational data extending close to or at their SED peaks, enabling reliable estimates of bolometric luminosity in these objects.

In many of the objects in our sample, it can be seen that some combination of host galaxy contribution and dust reradiation become significant redwards of \Ha . The hot toroidal dust component will be studied in greater detail in Paper II. Further discussion of the host contribution is presented in \S \ref{subsec:host}.

\section{Discussion}
\subsection{Model Suitability and Implications} \label{subsec:suitability}
We find that the \cite{done12} model is able to fit the IR to X-ray continuum of our sample of 11 $1.5 \lesssim z \lesssim 2.2$ AGN. We agree with the results of \cite{capellupo14} in that many of the objects can be modelled in the optical-IR regime by a geometrically thin, optically thick AD. In eight objects, constraints are put on the outer disc radius, which are compatible with considerations of the radius at which self-gravity truncates the AD \citep{laor89}. We note that the presence of a SX, observed and characterised by studies of local AGN, is both more physical and necessary to better define this continuum. The properties of this SX are related to the total energy of the Comptonised component, and modelling the X-ray spectrum in addition to optical/IR data is important to infer information about the SX.

Using our SED model we are able to place useful constraints on the bolometric luminosity for at least five of the objects in our sample. We believe that considerable uncertainties may arise if one assumes that the mass accretion rate is adequately estimated simply by the use of bolometric correction coefficients \citep{capellupo14}, as we infer a large spread in those parameters within our sample. In Table \ref{tab:sedprop} we show $\kappa_{\rm 2-10\,keV}$ and $\kappa_{5100 \rm \mathring{A}}$, two commonly-used proxies for the bolometric luminosity, for the sample. Though our sample is not large, we see a large range of values in all three models -- around a factor of 10 between the minimum and maximum. If we only consider the five objects with constrained SED peaks, this range is a factor of two in the $\kappa_{5100 \rm \mathring{A}}$, and a factor of 10 in $\kappa_{\rm 2-10\,keV}$, in spite of the  similar masses/accretion rates of these five AGN. This echoes the findings of \cite{elvis94} (who give the similar $L_{\rm bol} / L_{2500 \rm \mathring{A}}$ factor) and J12, and suggests that the spread is larger than the $\sim 20\%$ stated in \cite{capellupo14}. This cannot be solely due to the $\sim 0.1$ dex error on our mass estimate. We therefore suggest that BH spin is not the only property that cannot be estimated from singular properties of the optical spectra.

Our Model 3 provides the best fit to the data in all objects, judging from the \rchi \ fitting statistic, which takes into account the increased number of free parameters in Model 3 versus Models 1 and 2. In some objects the \rchi \ value is only marginally lower than the Model 2 value, which is indicative of a poorly constrained SX that does not benefit from the additional parameter freedom. Nonetheless, for the benefit of the objects in which the SX component is constrained (five objects), the additional freedom in Model 3 makes this the model of choice.

Many of the lower mass objects in our sample are predicted to have super-Eddington mass accretion rates, akin to the NLS1s (\eg J12a) and ULXs (\eg \citealt{sutton13}) we observe locally. However, we have not yet explored high-spin SED models in our study; this will be addressed in Paper II. It is possible that there may be some model degeneracy between spin and mass accretion rate, and that this contributes to the range of spins predicted by \citep{capellupo14} in their sample. We will therefore explore the effect on accretion rate of having higher spin BHs to assess whether the super-Eddington rates we have thus far predicted do indeed make these objects high mass NLS1 analogues, or whether it is more likely that the additional energy arises from moderately or highly spinning BHs.

Another limitation we have not yet explored is the reliability of our mass estimate. It is known that uncertainties on virial BH mass estimates are large ($\sim 0.1$ dex or greater) and in our study so far we have fixed it at the mean value. Allowing this to vary by 1--2 $\sigma$ may well improve the fit, or may add another source of degeneracy. Again, this will be explored in Paper II.

\subsection{Host Galaxy Contribution to the Optical/IR Continuum} \label{subsec:host}

Throughout this study we have made the assumption that any contribution to the SED from stars in the AGN host galaxy is likely to be negligible. This is a common assumption for typical quasars at $z>0.5$ (\eg \citealt{shen11}). However, we can test the validity of this assumption by superposing galaxy SED templates onto our faintest source, where the fractional stellar contribution will be largest. It is likely that the large galaxies that host the quasars in our sample are giant ellipticals, but it is known that starburst galaxies have significant energy output in the UV regime, and so we apply templates for both of these cases. 

We use two of the galaxy templates of \cite{polletta07} -- that of a 5 Gyr-old elliptical (appropriate for our redshift range) and that of the starburst galaxy M82, redshifted as appropriate. In terms of normalising these galaxy SEDs, we first assess the greatest possible contribution in J2328$+$1500 using the \M_BH -- $L_{\rm bulge}$ relation as presented in \eg \cite{marconi03} and \cite{degraf14}. We test J2328$+$1500 as it has the highest \M_BH, yet is our faintest source, and will thus almost certainly show the greatest contribution to the total SED by the host galaxy. The relations of both \cite{marconi03} and \cite{degraf14} predict a host galaxy of $M_V \simeq -25$.

\begin{figure}
\centering
\includegraphics[width = 8cm]{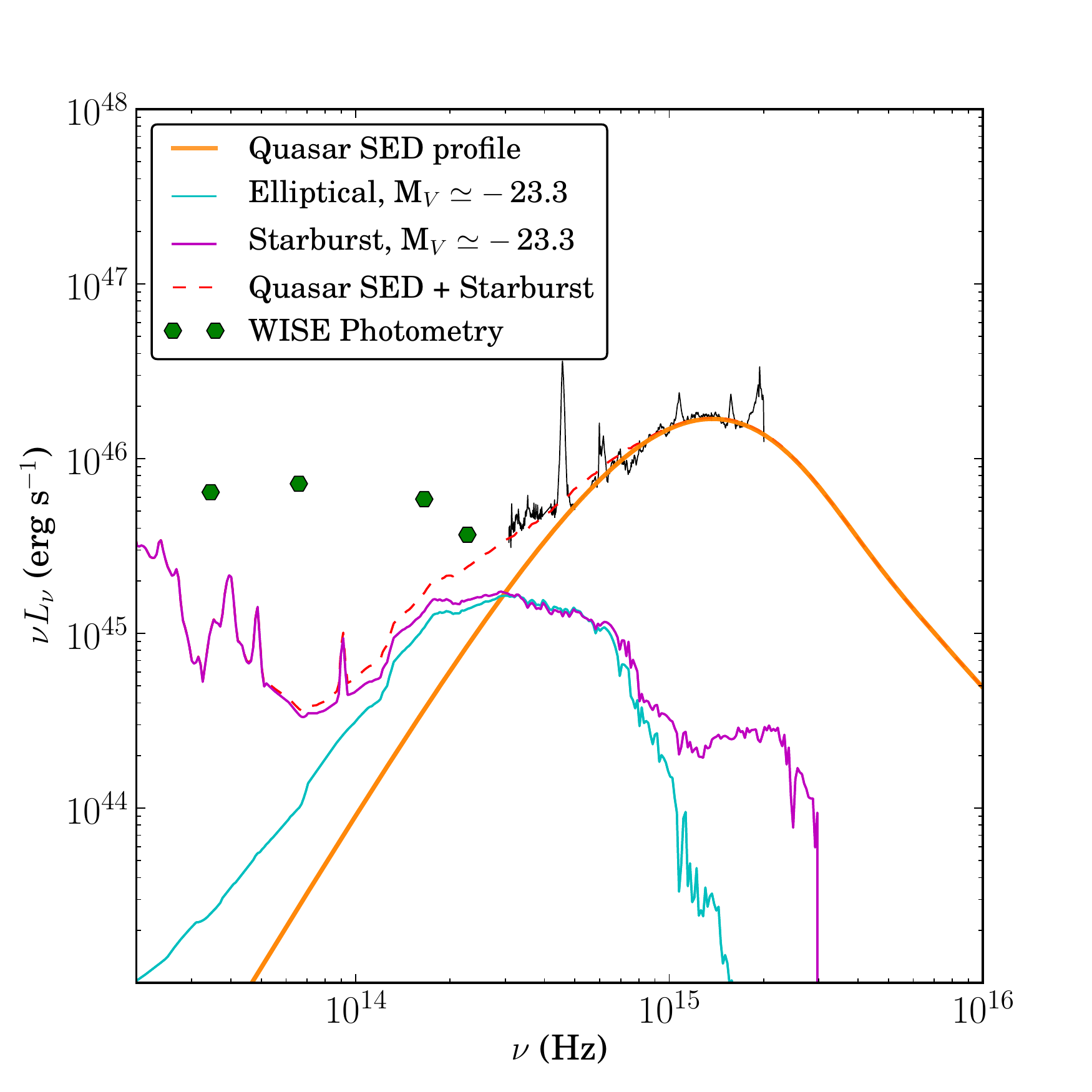}
\caption{\small Comparison of two host galaxy SED templates against the quasar SED for the faintest of our objects, J2328$+$1500. Host templates have been normalised to a V-band absolute magnitude of $M_V \simeq -23.3$, fainter than that predicted from the \M_BH -- $L_{\rm bulge}$ relation, but at the maximum possible contribution permitted by the data (dashed red line). Such a situation would imply some red contribution to the total flux of the source, but only a small contribution at the SED peak.}
\label{fig:host}
\end{figure}

We can put an upper limit on the host galaxy contribution using our SED model and the data. This greatest possible host contribution is shown in Fig.\ \ref{fig:host}, and corresponds to a host galaxy of $M_V \simeq -23.3$, around 1.7 magnitudes fainter than that predicted by the \M_BH -- $L_{\rm bulge}$ relation. If this were the case, the contribution to the host galaxy at the SED peak would be $\lesssim 2\%$, even for the case of a starburst galaxy (this template is $\sim 15$ times more luminous than M82). An elliptical host would make a negligible contribution at the SED peak.

This result, while representing an extreme case for this object, suggests that a host galaxy component may need to be included when we model the dusty torus component (evident in the WISE photometry) in Paper II. However, the contribution by the host galaxy to the total nuclear SED energy is small. For our other sources, which are brighter and ought to originate in smaller host galaxies (via the \M_BH -- $L_{\rm bulge}$ relation), the effect of the host will be smaller.

\subsection{Variability} \label{subsec:var}

AGN are known to exhibit variability across all wavelength ranges. Our study requires that the variability between the optical, IR and X-ray observations is not large. A discussion of our approach to detecting and correcting differences between the optical and IR spectral fluxes is given generally in \S \ref{subsec:dataprep}. To summarise, we only see evidence for a notable difference between optical and IR flux levels in J0041$-$0947. The origin of this change may be related to the Balmer continuum (as modelled by \citealt{shen12}), or poor quality photometry of 2MASS (to which the TSPEC spectrum was normalised).

Only a sub-set of properties of the AGN can change over timescales of a few years. The BH mass and spin are fixed, and changes in the mass accretion rate cannot occur faster than the viscous timescale, which is of the order of thousands of years. Another possible source of intrinsic variability in AGN may be tidal disruption events, in which tidal forces on a star passing close to the BH can produce large variations in the observed energy output.

Extrinsic effects can, in principle, modify the observed SED. Gravitational microlensing events by a star in a foreground galaxy may affect the total observed flux from an AGN, on timescales of months, though such events are predicted to be relatively rare. Current models of the torus suggest that it is likely to be clumpy, and so a change in the optical depth could occur if a clump were to drift into our line of sight. Indeed, major changes in the X-ray column density have been observed in several nearby AGN, on timescales of months to years (\eg \citealt{puccetti07}, \citealt{walton14}).

The properties of the intervening material in the AGN host galaxy will also have distinct modifying effects on the observed SED. In our SED models, we model attenuation as dust reddening of UV/optical/IR spectra, and photoelectric absorption of soft X-rays. This has the same wavelength dependence as that of the Milky Way, but is redshifted accordingly and has normalisation as a free parameter.

These effects will be explored in Paper II. To display the nature of variability in all our sources, we have plotted the available multi-epoch spectral data and photometry in the optical/IR bands for each object in Appendix \ref{app:dataplots}. This includes one epoch of photometry from each of the main surveys: SDSS photometric, UKIDSS and 2MASS, and all available epochs of spectral data. We supplement this with UV photometry from \xmm OM and GALEX all-sky imaging survey (AIS), to highlight the uncertainty in these values, which as mentioned in \S \ref{subsec:uv_xray} are unreliable due to absorption and the presence of emission features. Observation dates for the various data sources are tabulated in Appendix \ref{app:obsdates}. We have multi-epoch X-ray data for two of our objects -- J0839$+$5754 and J1044$-$2128. Treating each observation as a separate data set, we see no statistically significant evidence for variability in the X-ray spectra of these objects.

\subsection{Intrinsic Reddening} \label{subsec:intrinsicred}
\begin{figure}
\centering
\includegraphics[trim={0mm 0mm 0mm 0mm}, width = 8cm]{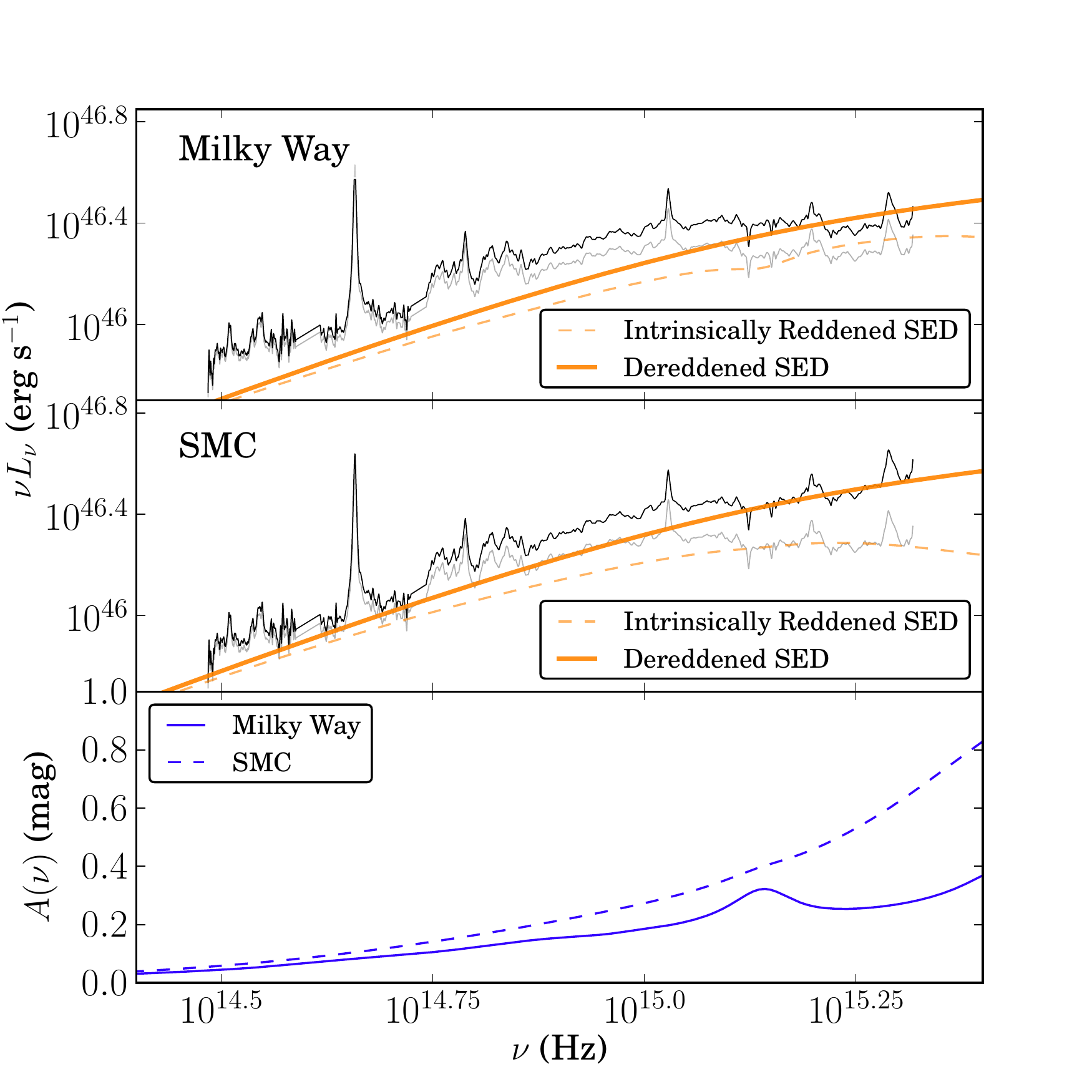}
\caption{\small Two different extinction curves -- the Milky Way and the SMC -- applied to J1044$+$2128. The Milky Way curve that we assume throughout this work clearly produces an inferior fit to the SMC reddening curve. The orange line shows the best-fit SED template, once corrected for intrinsic reddening, with the dotted orange line showing the intrinsically reddened SED. Similarly, in grey is the reddened optical/IR spectral data, and black the dereddened data. This data has been convolved with a 20-pixel Gaussian to smooth the data. The Milky Way model prioritises the higher S/N optical spectra, as these have a bigger effect on the \rchi \ fitting parameter than the noisier IR spectra.}
\label{fig:reddening_j1044}
\end{figure}

In our analysis, we have made the assumption that host galaxy dust extinction (intrinsic reddening) occurs via a similar process to extinction in the Milky Way. We thus use a redshifted Milky Way \cite{cardelli89} extinction curve, which produces apparently good reddening correction in all objects, except for J1044$+$2128. We have thus also tested two alternative models for dust extinction, those of the Large Magellanic Cloud (LMC) and the Small Magellanic Cloud (SMC). It is immediately apparent that a better continuum fit for J1044$+$2128 is achieved with the SMC extinction model, and this is corroborated by the \rchi \ fitting statistic. A comparison of different reddening curves is shown for this object in Fig.\ \ref{fig:reddening_j1044}. There is no evidence for the 2175 \AA \ feature in J1044$+$2128.

\begin{figure}
\centering
\includegraphics[trim={0mm 0mm 0mm 0mm}, width = 8cm]{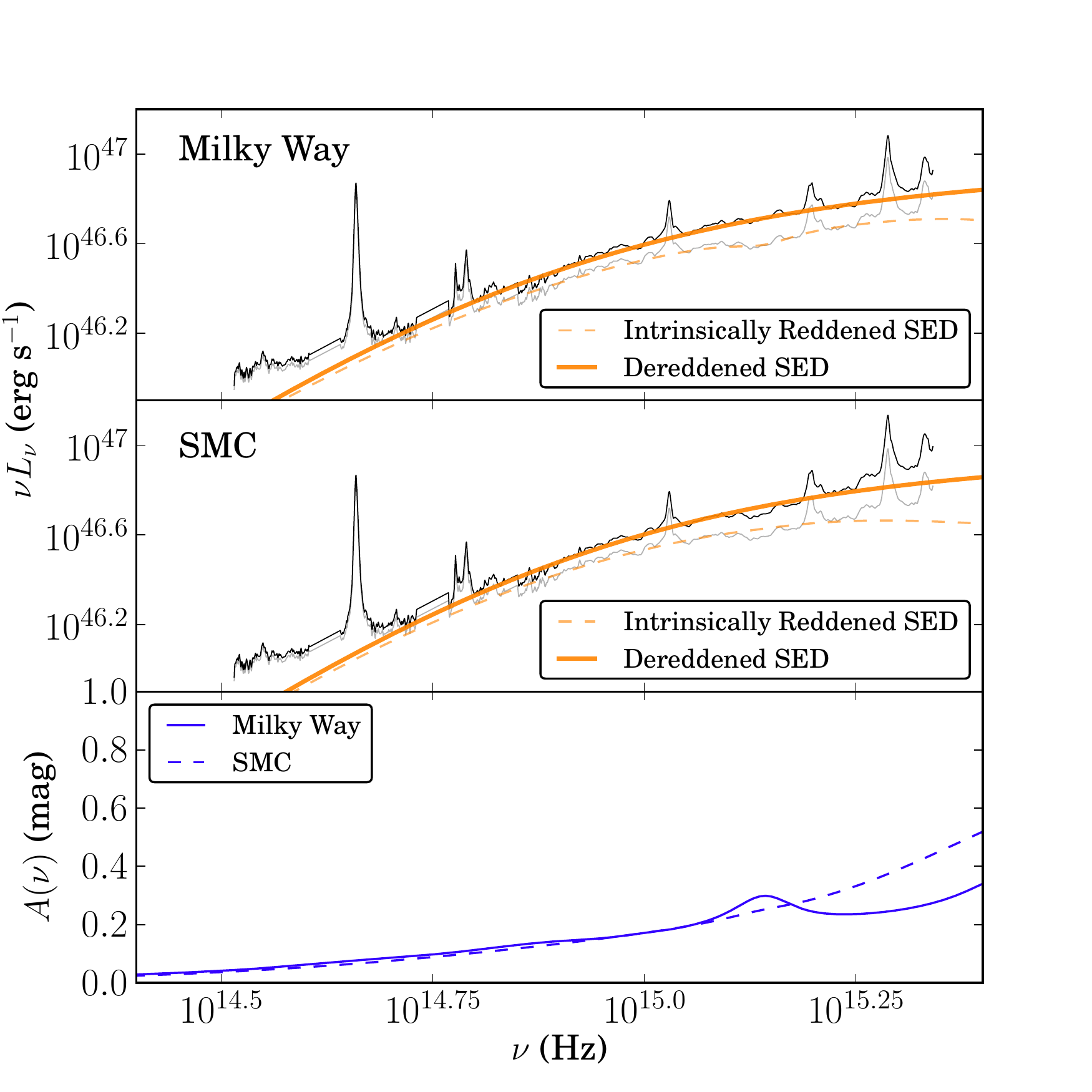}
\caption{\small An equivalent plot to Fig.\ \ref{fig:reddening_j1044} for J1350$+$2652. Here, to highlight the difference in continuum shapes implied by the two reddening curves, we have fixed both to the same $E(B-V)$ value (0.03 mag). In this object, the Milky Way reddening curve produces a noticeably better correction to the continuum shape.}
\label{fig:reddening_j1350}
\end{figure}

Judging by \rchi \ only, six objects are best fit with a Milky Way extinction curve (an example is given in Fig.\ \ref{fig:reddening_j1350}), one object with an SMC curve, and four objects with an LMC curve. In objects where the inferred intrinsic reddening is small, the difference between these $\chi^2$ values is marginal. \cite{capellupo14} came to a similar conclusion that different extinction curves are seen in different AGN, although they did not test the LMC model. We thus propose that Model 3 can be further augmented by including alternative reddening curves to the Cardelli curve used thus far. We will start Paper II by remodelling each SED with the best-fitting extinction curve.

Our only means of constraining the intrinsic reddening is the continuum shape, which is certainly a limitation -- we see in J2328$+$1500 that an E($B-V$) of less than 0.1 mag corresponds to an increase in \Lbol \ of 70\%. This is a limitation for all such studies. By investigating the effect of changing our model mass within the confidence limits of the mass estimate, as discussed in \S \ref{subsec:suitability}, we ought to be able to assess the objects in which changes to the SED slope due to reddening are degenerate with small changes in the mass estimate. Unfortunately these are most likely to occur in the objects with a sampled SED peak. Spin degeneracy may also prove to be a contributor. The best approach to test this is studying a larger sample in which correlations between, \eg intrinsic reddening and \Lbol \ are directly testable, which could help corroborate or rule out such degeneracies and sources of systematic errors.

\section{Summary and Conclusions}
In this study, we have selected a sample of 11 quasars at $1.5 \lesssim z \lesssim 2.2$. These objects all have optical, IR and X-ray spectral data, and UV photometry. We estimate the BH masses in each object using the \Ha line profile and method of \cite{greene05}, and then fit the  energy-conserving, three component SED model of \cite{done12} to each object. We fit three iterations of this model, adjusting the number of free parameters between each one. At this redshift range we would expect to observe the peak of the SED, due to the both the redshift, and the cooler accretion disc compared with AGN with lower mass BHs.

Our main conclusions are the following:

\begin{itemize}

\item[i.]{We observe the SED peak, or close to it, in five objects. We find that Model 3, which includes intrinsic attenuation and free SX normalisation fits best, allowing for the additional free parameters.}
\newline

\item[ii.]{When used in conjunction with the effects of dust reddening, we can accurately model the underlying optical-IR (rest UV-optical) continuum, and well-constrain the outer disc radius in eight objects.}
\newline

\item[iii.]{In the AGN with lower BH masses, we do not observe the SED peak, and in these cases the SX is therefore completely unconstrained. As a consequence of this, the model \rchi \ fitting parameters do not differ between the models with SX free and fixed for these objects. However, the SX contribution appears to be constrained to a varying degree in the five objects with data at the peak.}
\newline

\item[iv.]{Using template SEDs for both luminous elliptical galaxies and starbursts we show that the host galaxy contribution is insignificant at near to the peak of the SED, but it could contribute a fraction of the flux observed redward of \Ha . It is very likely that a dusty torus also contributes to flux here, judging from WISE photometry. We will therefore model this component accordingly in Paper II.}
\newline

\item[v.]{We show that UV photometry alone is insufficient to constrain the continuum. Ideally UV spectroscopy, \eg {\it HST}/COS, could be used to overcome the uncertainty of Ly-$\alpha$ forest absorption.}
\newline

\item[vi.]{The AGN in our sample generally have high Eddington ratios. In this respect they resemble the NLS1s, studied in nearby samples. This is expected, as in this redshift range, we preferentially observe the brightest AGN, which have high accretion rates. In Paper II we will test high-spin SED models, and there may be degeneracy in some of our objects between spin and mass accretion rate.}
\newline

\item[vii.]{We identify a range of properties in the best-fitting dust reddening component, with SMC/LMC reddening laws providing better fits than the Milky Way law in five objects.}
\newline

\item[viii.]{Our analysis provides more reliable estimates of the bolometric luminosity, as it uses data from across a large range of wavelengths, and utilises an energy-conserving SED model. We highight the problems of using a single parameter proxy, such as $\kappa_{2-10 \rm keV}$, as a means to derive \Lbol, as we see a large spread in such proxies, even in our small sample. We note that the six lowest mass objects have unsampled SED peaks, and therefore more poorly constrained bolometric luminosities. Having demonstrated the principle of applying our model successfully to multi-wavelength data, a much larger sample will be studied in Paper III to search for relationships between the overall SED characteristics and other specific emission line and continuum components.}

\end{itemize}

\section*{Acknowledgements}
The authors would like to thank an anonymous referee for constructive comments that improved the manuscript.

We are also very grateful to Yue Shen for kindly providing the TSPEC spectra for four of their objects.

We thank the contributors to the Python programming language, HEASARC, and IRAF, for software and services. We also thank the support from Gemini Observatory, who provided help on a number of occasions when the data was being reduced.

J.S.C. would like to acknowledge the support of an STFC studentship, and also useful advice from Charles Finn, Emma Gardner, Chichuan Jin and Sarah Hutton.

This research has made use of the VizieR catalogue access tool, CDS, Strasbourg, France. The original description of the VizieR service was published in \cite{ochsenbein00}. During this work, we made use of ``Ned Wright's Cosmology calculator'' \citep{wright06}.

\bibliography{Collinson_Quasar_SEDs_accepted}
\bibliographystyle{mn2e}

\newpage

\appendix

\clearpage

\section{SEDs, Model 1: SX fixed, no intrinsic attenuation} \label{app:sed1}
\noindent\begin{minipage}{\textwidth}
    \centering
    \includegraphics[trim={0mm 15mm 10mm 25mm}, clip, width = \textwidth]{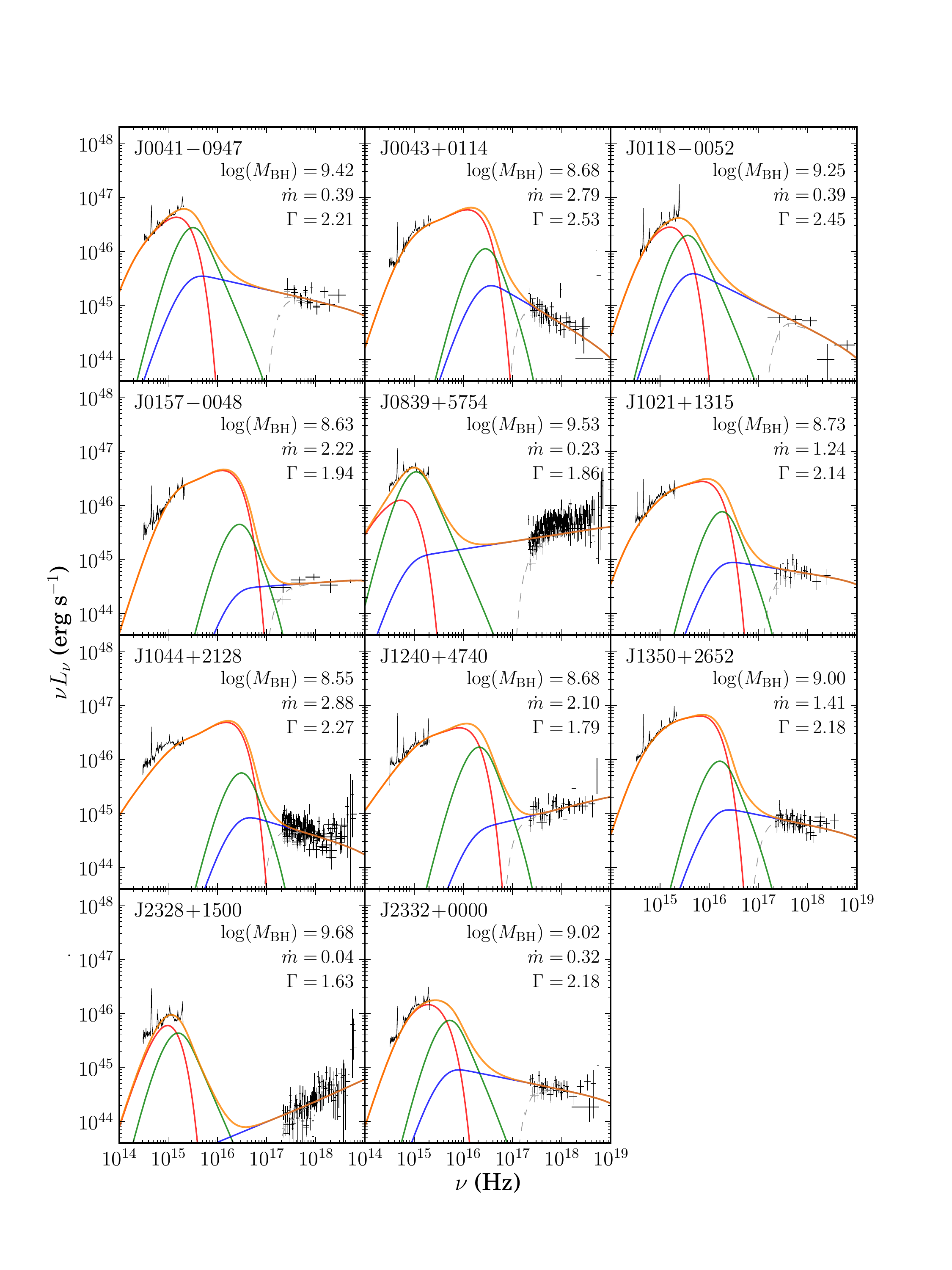}
    \captionof{figure}{\small Data and SED models for the sample. Here we use Model 1 (SX fixed, without intrinsic attenuation). We also plot the full IR-optical spectrum for each object. This spectral data is smoothed for clarity by convolving with a 20-pixel Gaussian. The different SED components are shown using the same colour scheme as in Fig.\ \ref{fig:sed}. The attenuated profile is shown by the dotted grey line.}
\label{fig:seds1}
\end{minipage}
\clearpage

\section{SEDs, Model 2: SX fixed, incl. intrinsic attenuation} \label{app:sed2}
\noindent\begin{minipage}{\textwidth}
    \centering
    \includegraphics[trim={0mm 15mm 10mm 25mm}, clip, width = \textwidth]{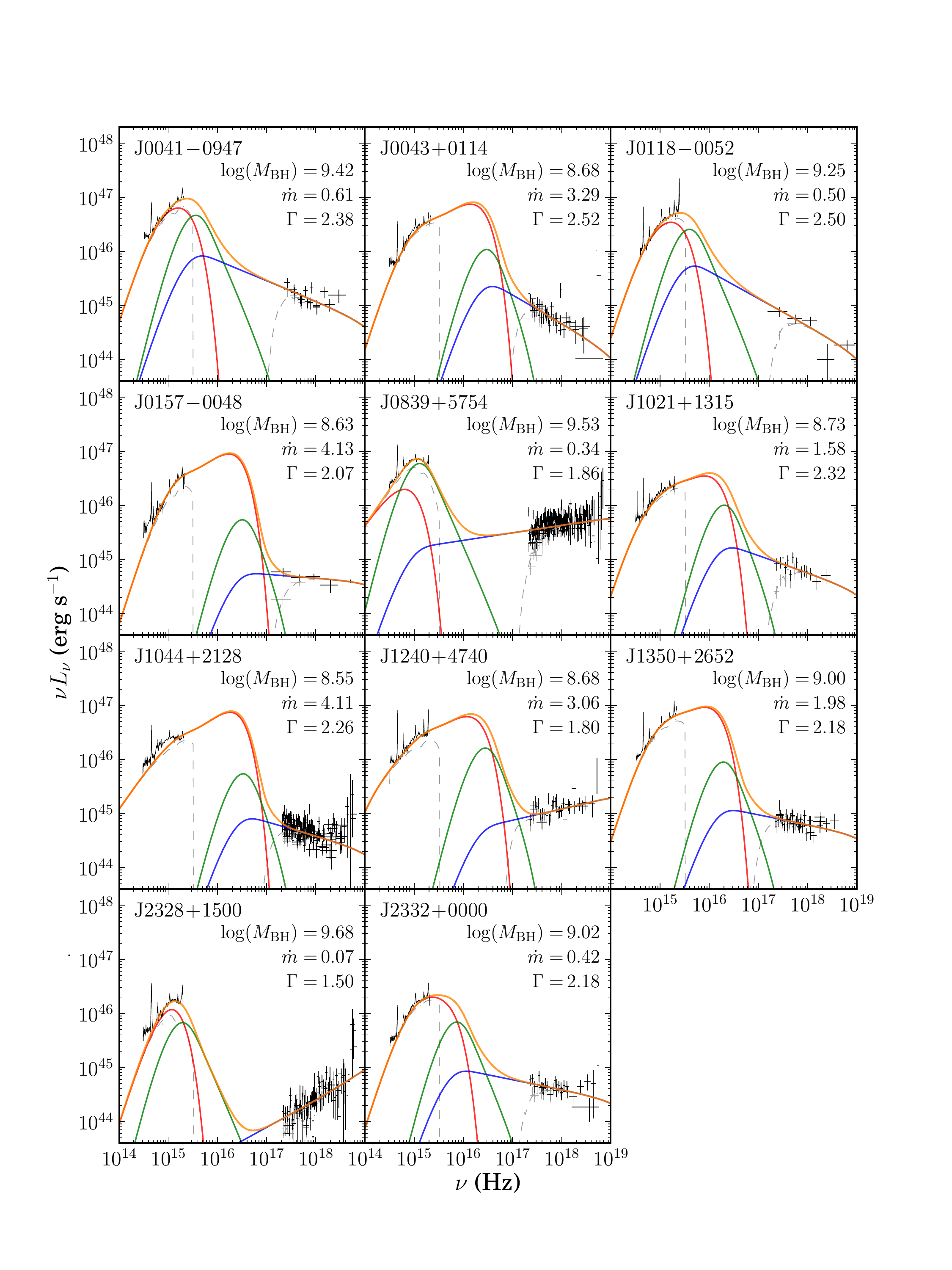}
    \captionof{figure}{\small Data and SED models for the sample. Here we use Model 2 (SX fixed, including intrinsic attenuation). The spectral data is smoothed as in Fig.\ \ref{fig:seds1}, and the same colour scheme is used.}
\label{fig:seds2}
\end{minipage}

\clearpage

\section{SEDs, Model 3: SX free, incl. intrinsic attenuation} \label{app:sed3}
\noindent\begin{minipage}{\textwidth}
    \centering
    \includegraphics[trim={0mm 15mm 10mm 25mm}, clip, width = \textwidth]{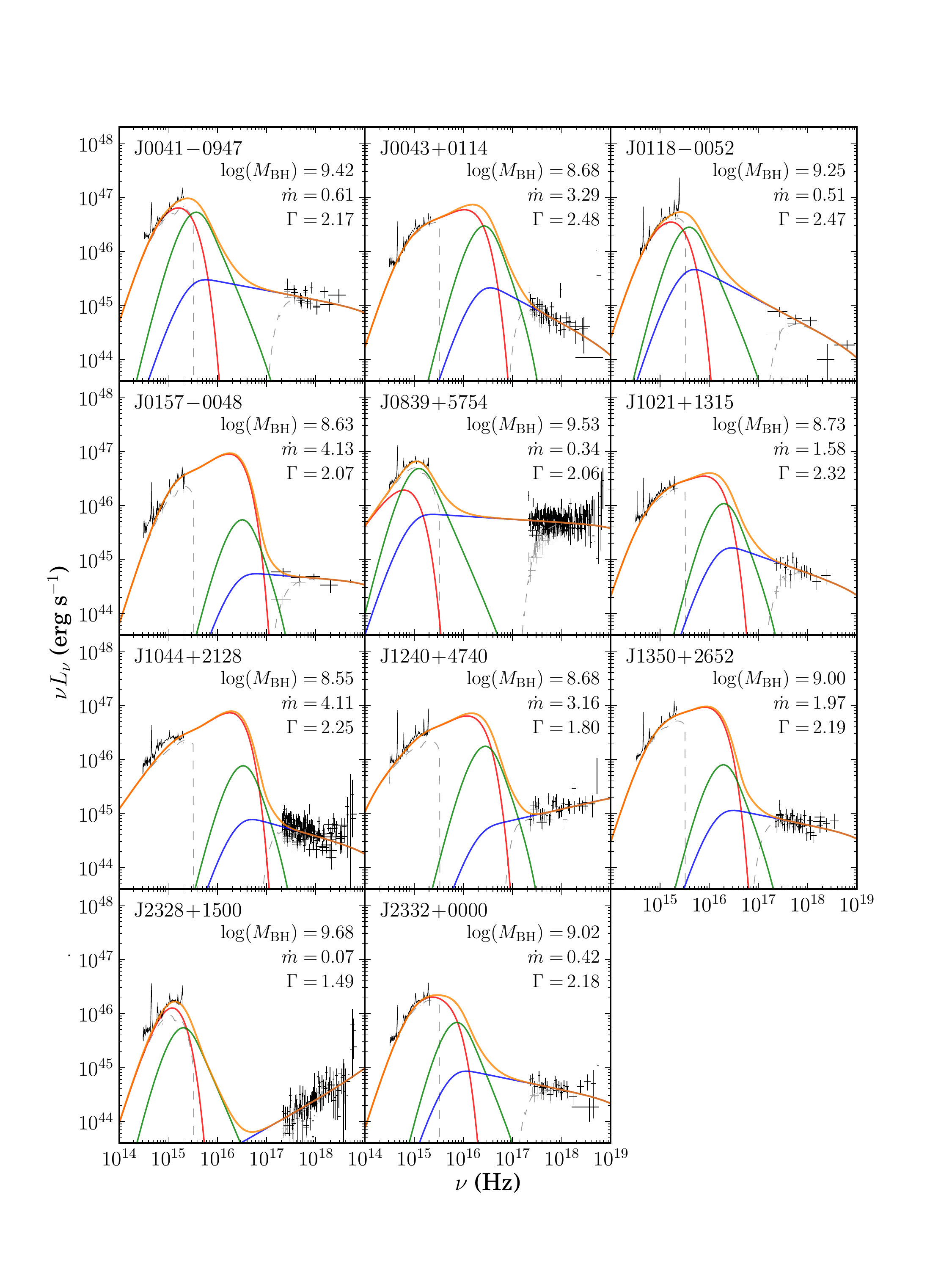}
    \captionof{figure}{\small Data and SED models for the sample. Here we use Model 3 (SX normalisation free, including intrinsic attenuation). The spectral data is smoothed as in Fig.\ \ref{fig:seds1}, and the same colour scheme is used.}
\label{fig:seds3}
\end{minipage}

\clearpage
\section{Spectral and Photometry Plots} \label{app:dataplots}
\noindent\begin{minipage}{\textwidth}
    \centering
    \includegraphics[trim={0mm 0mm 0mm 0mm}, clip, width = 0.92\textwidth]{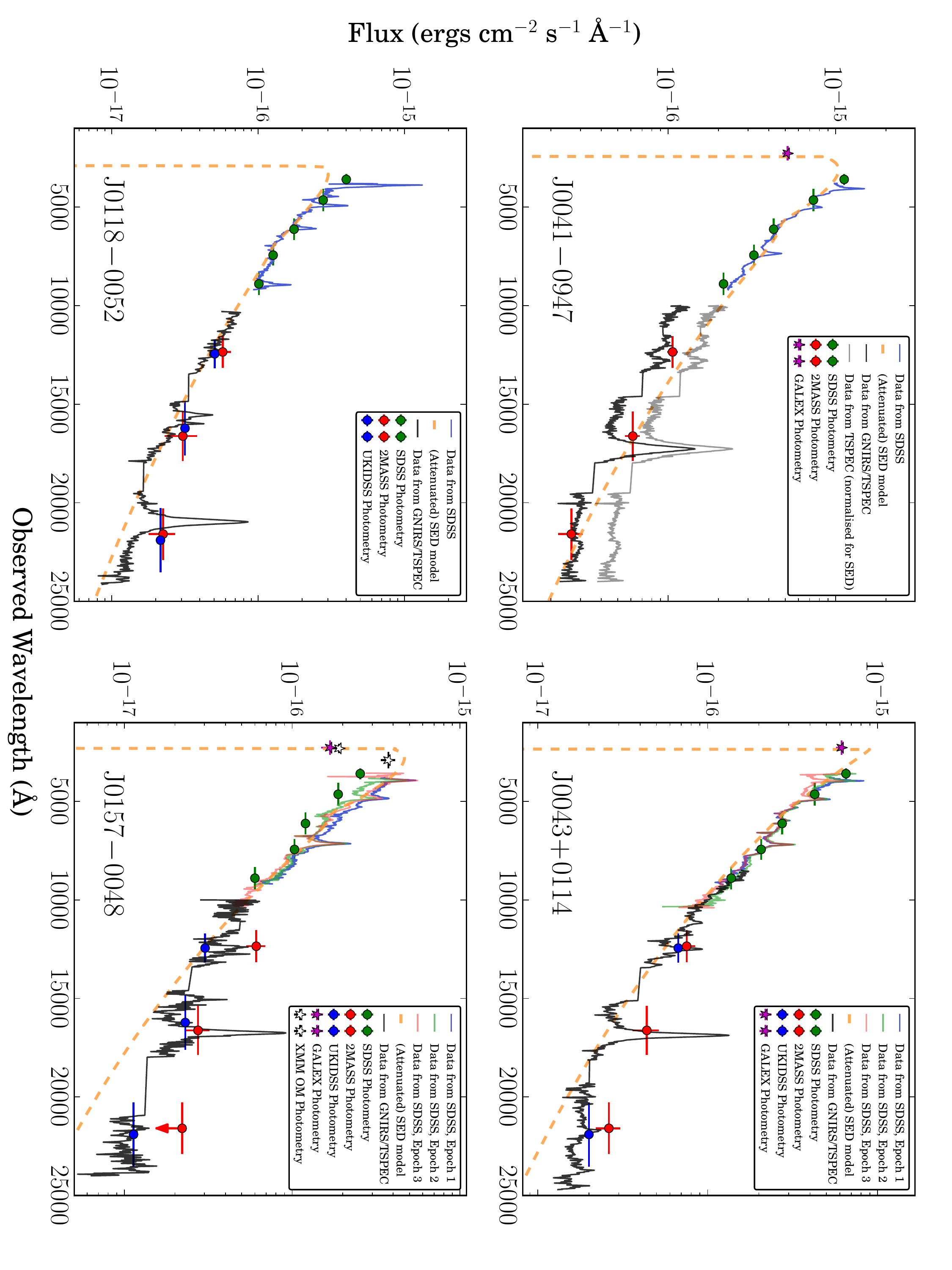}
    \captionof{figure}{\small All available spectral data plotted with photometry from large surveys. We also show the best fitting (Model 3) attenuated SED profile. At 912 \AA \ rest frame the photoelectric absorption component cuts the transmitted SED flux to zero. It can be seen in a few objects that have been observed on multiple occasions by SDSS/BOSS that variability or inconsistent flux calibration has occurred between observations. Observation dates are given in Appendix \ref{app:obsdates}.}
\label{fig:data1}
\end{minipage}

\clearpage

\noindent\begin{minipage}{\textwidth}
    \centering
    \includegraphics[trim={0mm 0mm 0mm 0mm}, clip, width = 0.92\textwidth]{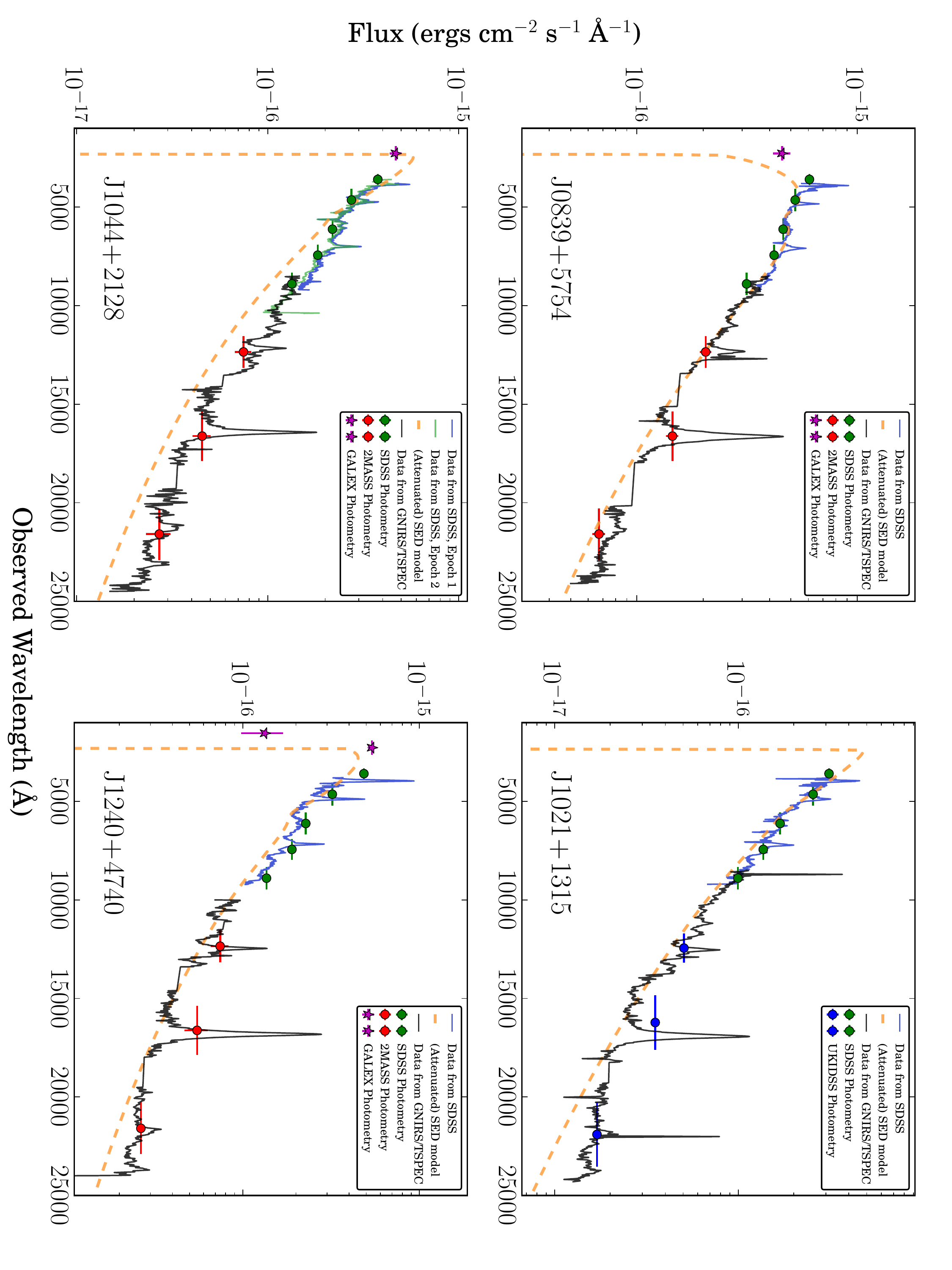}
    \captionof{figure}{\small Data plots continued.}
\label{fig:data2}
\end{minipage}

\clearpage

\noindent\begin{minipage}{\textwidth}
    \centering
    \includegraphics[trim={0mm 0mm 0mm 0mm}, clip, width = 0.92\textwidth]{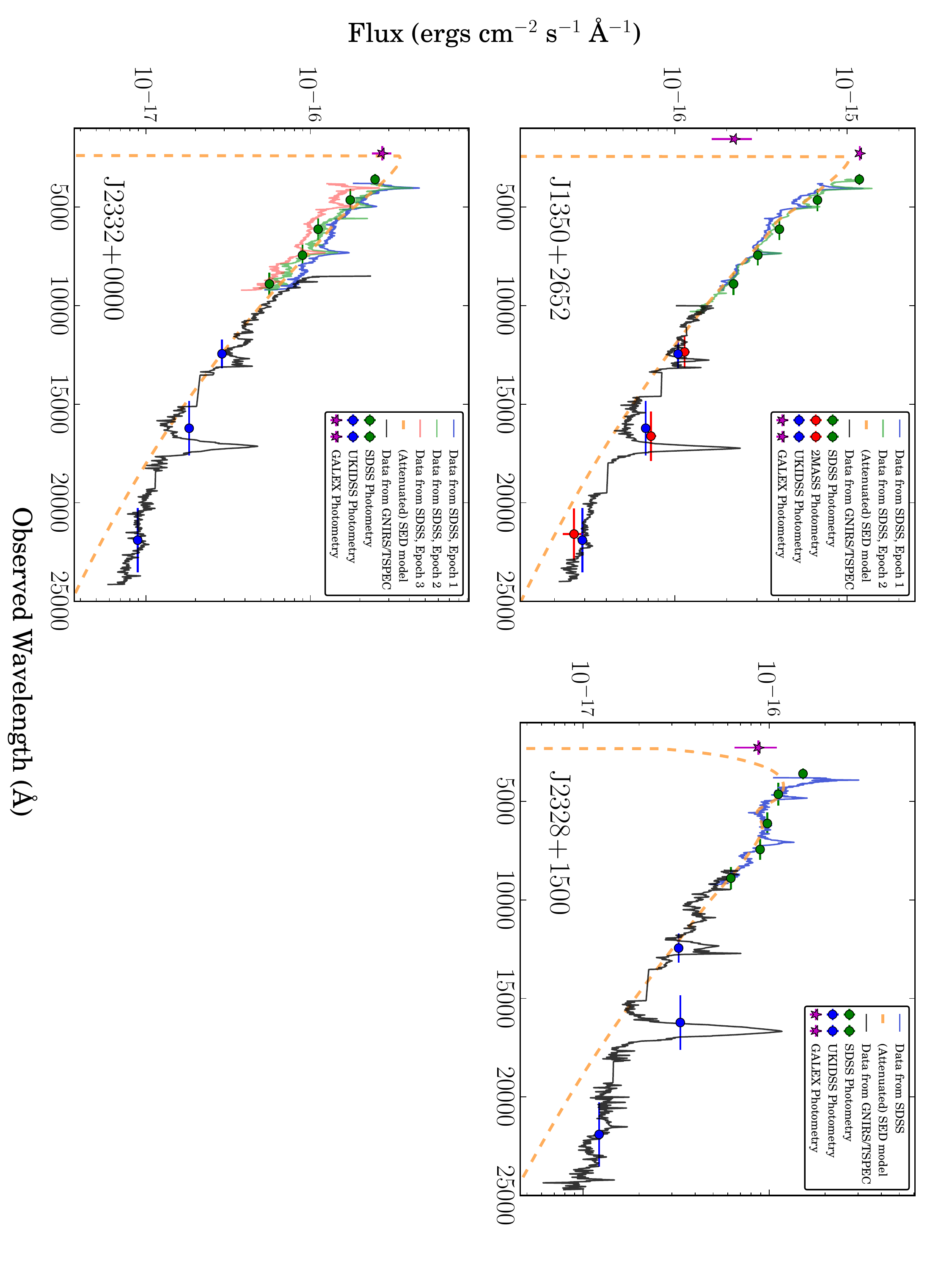}
    \captionof{figure}{\small Data plots continued.}
\label{fig:data3}
\end{minipage}

\clearpage

\section{Observation Dates} \label{app:obsdates}
\begin{table*}
\caption{\small Multi-epoch observation dates for all of the objects in our sample. We have searched all large area surveys offering good-quality data.}
\small
\centering
\begin{tabular}{c|ccc|ccc|ccc|ccc}
\hline
ID & \multicolumn{3}{c}{\bf{IR}}                  & \multicolumn{3}{c}{\bf{Optical}} 
   & \multicolumn{3}{c}{\bf{UV}}                  & \multicolumn{3}{c}{\bf{X-ray}}     \\
   
   & Instrument    & Type      & Date        & Instrument    & Type      & Date        
   & Instrument    & Type      & Date        & Instrument    & Type      & Date        \\

\hline
1  & 2MASS         & Phot      & 1998-10-02  & SDSS          & Phot      & 2000-09-25
   & GALEX         & Phot      & 2006-10-31  & \xmm EPIC     & Spec      & 2002-01-07   \\
   
   & TSPEC         & Spec      & 2010-01-02  & SDSS          & Spec      & 2001-09-10
   & GALEX         & Phot      & 2006-11-21  &               &           &              \\

   & TSPEC         & Spec      & 2010-11-28  &               &           &   
   & GALEX         & Phot      & 2011-03-18  &               &           &              \\
\hline
2  & 2MASS         & Phot      & 2000-11-29  & SDSS          & Spec      & 2000-09-07
   & GALEX         & Phot      & 2003-09-16  & \xmm EPIC     & Spec      & 2010-01-10   \\
   
   & UKIDSS        & Phot      & 2006-11-19  & SDSS          & Phot      & 2008-10-02
   & GALEX         & Phot      & 2003-09-30  &               &           &              \\

   & UKIDSS        & Phot      & 2008-11-28  & BOSS          & Spec      & 2009-12-21   
   & GALEX         & Phot      & 2007-10-27  &               &           &              \\
   
   & GNIRS         & Spec      & 2013-08-16  & BOSS          & Spec      & 2010-09-05
   &               &           &             &               &           &              \\   
\hline
3  & 2MASS         & Phot      & 1998-09-18  & SDSS          & Spec      & 2000-09-02
   & GALEX         & Phot      & 2008-10-20  & \xmm EPIC     & Spec      & 2003-07-11   \\
   
   & GNIRS         & Spec      & 2004-11-29  & SDSS          & Phot      & 2004-09-23
   & GALEX         & Phot      & 2008-10-31  &               &           &              \\

   & UKIDSS        & Phot      & 2006-07-10  &               &           &              
   & GALEX         & Phot      & 2008-11-16  &               &           &              \\
   
   &               &           &             &               &           &           
   & GALEX         & Phot      & 2011-10-28  &               &           &              \\
\hline
4  & 2MASS         & Phot      & 1998-09-29  & SDSS          & Spec      & 2000-11-23
   & GALEX         & Phot      & 2004-10-11  & \xmm EPIC     & Spec      & 2005-07-14   \\
   
   & UKIDSS        & Phot      & 2005-09-07  & SDSS          & Spec      & 2001-09-27
   & \xmm OM       & Phot      & 2005-07-14  &               &           &              \\

   & TSPEC         & Spec      & 2009-11-07  & SDSS          & Phot      & 2003-11-19   
   & GALEX         & Phot      & 2008-10-20  &               &           &              \\
   
   & TSPEC         & Spec      & 2010-11-28  & BOSS          & Spec      & 2010-09-10 
   &               &           &             &               &           &              \\
\hline
5  & 2MASS         & Phot      & 2000-01-05  & SDSS          & Phot      & 2003-10-23
   & \xmm OM       & Phot      & 2006-10-03  & \xmm EPIC     & Spec      & 2006-10-03   \\
   
   & GNIRS         & Spec      & 2013-10-27  & SDSS          & Spec      & 2007-11-21
   & GALEX         & Phot      & 2007-01-03  & \xmm EPIC     & Spec      & 2007-04-06   \\

   &               &           &             &               &           &              
   & \xmm OM       & Phot      & 2007-05-09  & \xmm EPIC     & Spec      & 2007-05-09   \\
   
   &               &           &             &               &           &            
   & GALEX         & Phot      & 2010-01-14  &               &           &              \\
\hline
6  & UKIDSS        & Phot      & 2010-02-08  & SDSS          & Phot      & 2003-01-27
   & \xmm OM       & Phot      & 2003-05-05  & \xmm EPIC     & Spec      & 2003-05-05   \\
   
   & GNIRS         & Spec      & 2014-03-21  & SDSS          & Spec      & 2004-02-27
   & GALEX         & Phot      & 2006-03-27  &               &           &              \\

   &               &           &             &               &           &   
   & GALEX         & Phot      & 2010-03-14  &               &           &              \\
\hline
7  & 2MASS         & Phot      & 1998-01-29  & SDSS          & Phot      & 2005-03-09
   & GALEX         & Phot      & 2006-09-14  & \xmm EPIC     & Spec      & 2003-05-05   \\
   
   & GNIRS         & Spec      & 2014-03-20  & SDSS          & Spec      & 2006-12-28
   &               &           &             & \xmm EPIC     & Spec      & 2003-05-28   \\

   &               &           &             & BOSS          & Spec      & 2012-04-22  
   &               &           &             & \xmm EPIC     & Spec      & 2003-12-12   \\
\hline
8  & 2MASS         & Phot      & 1998-05-16  & SDSS          & Phot      & 2003-03-10
   & \xmm OM       & Phot      & 2002-11-12  & \xmm EPIC     & Spec      & 2002-11-12   \\
   
   & TSPEC         & Spec      & 2011-02-22  & SDSS          & Spec      & 2004-03-25
   & GALEX         & Phot      & 2007-03-04  &               &           &              \\
\hline
9  & 2MASS         & Phot      & 2000-04-11  & SDSS          & Phot      & 2004-06-11
   & \xmm OM       & Phot      & 2004-01-25  & \xmm EPIC     & Spec      & 2004-01-25   \\
   
   & UKIDSS        & Phot      & 2010-03-01  & SDSS          & Spec      & 2006-04-23
   & GALEX         & Phot      & 2006-04-30  &               &           &              \\

   & TSPEC         & Spec      & 2011-02-22  & BOSS          & Spec      & 2012-06-27  
   & GALEX         & Phot      & 2009-05-27  &               &           &              \\
   
   &               &           &             &               &           &            
   & GALEX         & Phot      & 2011-05-27  &               &           &              \\
\hline
10 & UKIDSS        & Phot      & 2007-09-28  & SDSS          & Phot      & 2000-09-26
   & GALEX         & Phot      & 2004-09-15  & \xmm EPIC     & Spec      & 2007-12-01   \\
   
   & GNIRS         & Spec      & 2013-08-18  & SDSS          & Spec      & 2001-11-25
   & GALEX         & Phot      & 2006-02-26  &               &           &              \\

   &               &           &             &               &           &        
   & GALEX         & Phot      & 2007-03-28  &               &           &              \\
   
   &               &           &             &               &           &            
   & GALEX         & Phot      & 2009-09-09  &               &           &              \\

   &               &           &             &               &           &            
   & GALEX         & Phot      & 2009-10-07  &               &           &              \\
\hline
11 & UKIDSS        & Phot      & 2006-05-06  & SDSS          & Spec      & 2000-10-04
   & GALEX         & Phot      & 2004-03-14  & \xmm EPIC     & Spec      & 2007-12-01   \\
   
   & GNIRS         & Spec      & 2013-08-19  & SDSS          & Spec      & 2001-10-17
   & GALEX         & Phot      & 2006-10-01  &               &           &              \\

   &               &           &             & SDSS          & Spec      & 2002-09-08 
   & GALEX         & Phot      & Many obs    &               &           &              \\
   
   &               &           &             & SDSS          & Phot      & 2003-11-19 
   &               &           & in DIS      &               &           &              \\
\hline

\end{tabular}
\label{tab:epochs}
\end{table*}

\label{lastpage}

\end{document}